\documentclass[final,1p,times]{elsarticle}

\usepackage{amsmath,amsthm,amssymb,bm}

\usepackage{lineno}

\usepackage{epsfig}
\usepackage{amsfonts}
\usepackage{algorithm}
\usepackage{algorithmic}
\usepackage{natbib}
\usepackage{color,colortbl}

\newcommand{\be}{\begin{equation}}
\newcommand{\ee}{\end{equation}}
\newcommand{\bea}{\begin{eqnarray}}
\newcommand{\eea}{\end{eqnarray}}
\newcommand{\nn}{\nonumber}

\newcommand{\vect}[1]{\boldsymbol{#1}}
\newcommand{\mathtiny}[1]{\mbox{\tiny $#1$}}

\definecolor{LightCyan}{rgb}{0.88,1,1}
\definecolor{Violet}{rgb}{0.93,0.51,0.93}

\biboptions{comma,round}

\journal{Computational Statistics \& Data Analysis}

\begin{document}

\begin{frontmatter}



\title{Systematic Physics Constrained Parameter Estimation of Stochastic
  Differential Equations}

\author[label1]{Daniel Peavoy\fnref{label4}}
\address[label1]{Complexity Science Centre, University of Warwick, Coventry, UK}
\author[label2]{Christian L. E. Franzke\corref{cor1}}
\address[label2]{Meteorological Institute and Center for Earth System Research
  and Sustainability (CEN), University of Hamburg, Hamburg, Germany}
\author[label3]{Gareth O. Roberts}
\address[label3]{Department of Statistics, University of Warwick, Coventry, UK}

\cortext[cor1]{Center for Earth System Research and Sustainability
  (CEN), University of Hamburg, Grindelberg 7, 20144 Hamburg, Germany,
  christian.franzke@gmail.com}

\fntext[label4]{Present affiliation: Manufacturing Technology Centre,
  Coventry, UK}



\begin{abstract}
A systematic Bayesian framework is developed for physics constrained parameter
inference of stochastic differential equations (SDE) from partial
observations. Physical constraints are derived for stochastic climate
models but are applicable for many fluid systems. A condition is derived for
global stability of stochastic climate models based on energy
conservation. Stochastic climate models are globally stable when a quadratic
form, which is related to the cubic nonlinear operator, is negative
definite. A new algorithm for the efficient sampling of such negative definite
matrices is developed and also for imputing unobserved data which improve the
accuracy of the parameter estimates. The performance of this framework is
evaluated on two conceptual climate models.
\end{abstract}


\end{frontmatter}


\section{Introduction}
\label{Sec:1}
In many areas of science the inference of reduced order hybrid
dynamic-stochastic models, which take the form of stochastic differential
equations (SDE), from data is very important. For many applications
running full resolution dynamical models is computationally prohibitive and in
many situations one is mainly interested in large-scale features and not the
exact evolution of the fast, small scale features, which typically determine
the time step size. Thus, reduced order stochastic models are an attractive
alternative. Examples are molecular dynamics \citep{Horenko:2005}, engineering
turbulence \citep{Heinz:2003} and climate science
\citep{Franzke:2005,Franzke:2006,Kondrashov:2006}.

The inference of such models has been done using non-parametric methods
\citep{Siegert:1998,Friedrich:2000,Crommelin:2006} from partial
observations. These non-parametric methods need very long time series for
reliable parameter estimates and can be used only for very low-dimensional
models because of the curse of dimension. More importantly, they do not
necessarily obey conservation laws or stability properties of the full
dimensional dynamical system. In many areas of science one can derive reduced
order models from first principles \citep{Majda:2009,Pavliotis:2008} such that
certain fundamental properties of the full dynamics are still valid. These
methods provide us with parametric forms for the model fitting. Physical
constraints then not only constrain the parameters one has to estimate but they
can also ensure global stability. Thus, there is a need for systematic physics
constrained model and parameter estimation procedures
\citep{Peavoy:2013,Majda:2013}.

For instance, the climate system is governed by conservation laws like energy
conservation. Based on this energy conservation property the normal form of
stochastic climate models has been derived by \cite{Majda:2009} using the
stochastic mode reduction procedure
\citep{Majda:1999,Majda:2001,Majda:2008,Franzke:2005,Franzke:2006}. This
procedure allows the systematic derivation of reduced order models from first
principles. This normal form provides a parametric form for parameter
estimation from partial observations which we will use in this study.

The fundamental form of climate models is given by
\begin{equation}
\frac {d \mathbf{z}} {dt} = F + L \mathbf{z} + B(\mathbf{z},\mathbf{z}),
\end{equation}
where $ \mathbf{z} \in \mathbb{R}^N $ denotes the N-dimensional state vector,
F the external forcing, L a linear and B a quadratic nonlinear operator. The
nonlinear operator B is conserving energy $ \mathbf{z} \cdot
B(\mathbf{z},\mathbf{z}) $. For current climate models N is of the
order of $ 10^6 - 10^8 $. This shows that running complex climate models is
computationally expensive. But for many applications like extended-range
(periods of more than 2 weeks), seasonal and decadal climate predictions one
is only interested in the large-scale circulation of the climate system and
not whether there will be a cyclone over London on a particular
day next year. The large-scale circulation can successfully be predicted using
reduced order models
\citep{Selten:1995,Achatz:1999,Franzke:2005,Franzke:2006,Kondrashov:2006}.

The stochastic mode reduction procedure
\citep{Majda:1999,Majda:2001,Majda:2008} provides a systematic framework for
deriving reduced order climate models with a closure which takes account of
the impact of the unresolved modes on the resolved modes. In order to derive
reduced order models one splits the state vector $ \mathbf{z} = \left(
\begin{array}{c} \mathbf{x} \\ \mathbf{y} \end{array} \right) $ into resolved 
$ \mathbf{x} $ and unresolved $ \mathbf{y} $ modes. The stochastic mode
reduction procedure now enables us to systematically derive a reduced order
climate model which only depends on $ \mathbf{x} $
\begin{equation}
d \mathbf{x} = \left( \tilde{F} + \tilde{L} \mathbf{x} +
\tilde{B}(\mathbf{x},\mathbf{x}) + M(\mathbf{x},\mathbf{x},\mathbf{x}) \right)
dt + a(\mathbf{x},\bm{\sigma}) d\mathbf{W},
\label{EQ:SDE}
\end{equation}
where $M$ denotes a cubic nonlinear term, $ W $ is the Wiener process
and $\sigma$ the diffusion parameters.

In this study we will develop a systematic Bayesian framework for the
efficient estimation of the model parameters using Markov Chain Monte Carlo
(MCMC) methods from partial observations. We are dealing with partial
observations because we now only have knowledge of the few resolved
modes $ \mathbf{x} $ and are ignorant about the many unresolved modes
$ \mathbf{y} $. 

Stochastic climate modeling is a complex problem and empirically estimating
the parameters poses several problems. First, the nonlinearity of climate models
requires an approximation of the likelihood function. While it can be shown
that this approximation converges to the true likelihood, this is not
necessarily the case for real world applications. Here we develop a MCMC
algorithm for the first time for stochastic climate models and demonstrate
that this algorithm performs well. Second, the nonlinearity
of the problem causes the space of parameters leading to stable and physical
meaningful solutions to become small as the dimension of the problem
increases. We show that a lot of the posterior mass is on parameter
values which lead to solutions exploding to infinity in finite time. To solve
this problem we derive global stability conditions. These conditions take the
form of a negative definite matrix. Hence, we devise a novel sampling
strategy based on sampling non-negative matrices. We show that this sampling
strategy is computationally efficient and leads to stable solutions.

In section \ref{Sec:2} we introduce stochastic climate models and
derive conditions for global stability. Previous studies have shown that
reduced climate models with quadratic nonlinearity experience unphysical finite
time blow up and long time instabilities
\citep{Harlim:2014,Majda:2012,Majda:2013,Yuan:2011}. Here we use the normal
form of stochastic climate models \citep{Majda:2009} and derive sufficient
conditions for global stability for the normal form of stochastic climate
models. These normal form stochastic climate models have cubic
nonlinearities. The here derived stability condition is more general than the
one in \citet{Majda:2009}. In section \ref{Sec:3} we develop a Bayesian
framework for the systematic estimation of the model parameters using physical
constraints. Here we develop an efficient way of sampling negative-definite
matrices. Without this constraint the MCMC algorithm would produce about 40\%
unphysical solutions which is clearly very inefficient. Here we also
demonstrate that for these kinds of SDEs  imputing data improves the
parameter estimates considerably. In section \ref{Sec:4} we demonstrate the
accuracy of our framework on conceptual climate models. We summarize our
results in section \ref{Sec:6}.

\section{Stochastic Climate Models and Global Stability}
\label{Sec:2}
Here we study the following D dimensional normal form of stochastic
climate models (which has the same structural form as
Eq. (\ref{EQ:SDE}), see \citep{Majda:2009}):
\begin{subequations}
\begin{eqnarray}
dx_i & = & \left( \alpha_i + \sum_{j=1}^D \beta_{i,j} x_j + \sum_{j+1}^D
\sum_{k=1}^j \gamma_{i,j,k} x_j x_k + \sum_{j=1}^D \sum_{k=1}^j \sum_{l=1}^k
\lambda_{i,j,k,l} x_j x_k x_l \right) dt \\
& + & \sum_{j=1}^D a_{i,j} dW_j + \sum_{j=1}^D \sum_{k=1}^j b_{i,j,k} x_j dW_k.
\end{eqnarray}
\label{EQ:0}
\end{subequations}
which we write for convenience in a more compact form
\begin{equation}
d \mathbf{x} = \mu(\mathbf{x},\mathbf{A})dt + {\bf a}(\mathbf{x},\bm{\sigma})dW.
\label{EQ:1}
\end{equation}

The parameters $ \alpha, \beta, \gamma$ and $\lambda$ are written as one
matrix $\mathbf{A} \in \mathbb{R}^{D \times P}$. We allow for the inclusion of all
possible linear, quadratic and cubic terms with forcing terms entering first,
followed by linear, quadratic and then the cubic terms. We include them into a
matrix $\mathbf{A}$ as $ A_{i,1} = \alpha_i $, $ A_{i,j+1} = \beta_{i,j}$,
$A_{i,j(j-1)/2+k+D+1} = \gamma_{i,j,k} $ and $ A_{i,f(j,k,l)} =
\lambda_{i,j,k,l} $. The index function for the cubic term is given by
\begin{equation}
f(j,k,l) = 1 + D + \frac {D(D+1)} {2} + \frac {j(j-1)(j+1)} {6} + \frac
{k(k-1)} {2} + l.
\end{equation}

Global stability implies that the cubic terms act as a nonlinear damping. For
climate models energy conservation of the nonlinear operator implies global
stability. However, in general, global stability does not necessarily imply
energy conservation. The energy equation based only on the cubic terms can be
written as 
\begin{equation}
\frac {1} {2} \frac {dE} {dt} = \sum_{i=1}^D x_i \frac {dx_i} {dt} =
\sum_{i=1}^D \sum_{j=1}^D \sum_{k=1}^j \sum_{l=1}^k A_{i,f(j,k,l)} x_i x_j x_k x_l.
\end{equation}
Here we only consider the cubic term because this term will ultimately
determine global stability. \citet{Majda:2009} have shown that the normal form
of stochastic climate models allows for linearly unstable modes. These
linearly unstable modes are associated with important weather systems and
waves and are an intrinsic and important part of climate models. Once these
linearly unstable modes have reached a certain amplitude the nonlinear cubic terms
will govern their evolution and, thus, ensure global stability.

We consider now the vector $ \mathbf{v} $ with $ \frac {D(D+1)} {2} $ components of the
form $ v_{(i-1)i/2+j} = x_i x_j $ with $ 1 \le j \le i \le D $. Now we find a
negative definite matrix $\mathbf{M}$ such that \cite{Majda:2009}
\begin{equation}
\mathbf{v}^T \mathbf{M} \mathbf{v} = \frac {1} {2} \frac {dE} {dt} \leq 0.
\end{equation}
A sufficient solution is as follows: Let matrix $M \in \mathbb{R}^{(D+1)D/2
\times (D+1)D/2}$ be
\begin{equation}
M_{(i-1)i/2+j,(k-1)k/2+l} = \left\{ \begin{array}{l l} 
A_{i,f(i,j,l)}, & \mbox{ if k$>$j and l $\le$ j} \\
0, & \mbox{ if k$>$ j and l$>$ j} \\
A_{i,f(i,j,l)} + A_{i,f(i,j,k)}, & \mbox{ if k $\le$ j and  l$<$k} \\
A_{i,f(i,j,l)}, & \mbox{ if k $\le$ j and l$=$k},
\end{array} \right.
\end{equation}
where $ 1 \le j \le i \le D$ and $ 1 \le l \le k \le D $. While this solution
is not necessarily unique the imposition of this constraint is still necessary
in order to reduce the amount of parameter values leading to unstable models
and, thus, to reduce the computational expense of the parameter inference. As
we will show below, not imposing this constraint will lead to unstable and
unphysical solutions in about 40\% of parameter values in our MCMC scheme.

In summary, the stochastic climate model in Eq. (\ref{EQ:1}) is globally
stable if the tensor $\mathbf{M}$ is negative definite and $M$ determines the
components of the cubic operator $\lambda$. This is an important result for
the constrained parameter estimation of stochastic climate models
\citep{Majda:2009,Peavoy:2013}.
 
\section{Physics Constraint Parameter Sampling}
 \label{Sec:3}
We use a Markov Chain Monte Carlo algorithm for the parameter inference which
was proposed by \cite{Chib:2004} and \cite{Golightly:2008}. We use this
approach because of its flexibility. For instance, the exact algorithms by
\cite{Beskos:2005} and \cite{Beskos:2008} cannot be easily applied to
multidimensional diffusions. Furthermore, the exact algorithms require that
the drift function must be the gradient of a potential; for instance,
stochastic climate models cannot be written in such a form. 

Our MCMC algorithm first updates the diffusion parameters (see
algorithm 1), then updates the imputed data (algorithm 2) and finally
updates the drift parameters (section 3.2). To efficiently propose
imputed data we us the Modified Linear Bridge sampler (section
3.1). To physically constrain the drift parameters we develop a
scheme to sample negative definite matrices (section 3.3 and algorithms
3 and 4).

The novel aspect of our MCMC algorithm is the physics constraint
sampling which ensures the stability of the reduced stochastic model. Moreover,
this algorithm overcomes the dependency between the diffusion parameters and
the missing data by changing variables to the underlying Brownian motion $
\vect{W} \in \mathbb{R}^d $ and conditioning on this when performing the
parameter update. This ensures consistency between the parameters and the path
\cite{Chib:2004,Golightly:2008}. In order to improve the accuracy of the Euler
approximation we introduce latent data points between all pairs of
observations. While this is not trivial for nonlinear models this can be
accomplished by introducing a suitable diffusion bridge
\cite{Durham:2002,Golightly:2008}. For this purpose we define a process
$\vect{Z}$, which conditions on the endpoint $\vect{x}_T$, by
\be
d\vect{X}_t=\vect{a}(\vect{X}_t,\vect{\sigma})d\vect{Z}_t+\frac{\vect{x}_T-\vect{X}_t}{T-t}dt,\,\,\vect{X}_0=\vect{x}_0\,,
\ee
where $T$ is the next observation time. In discrete time the transformation is
\be 
\vect{X}_{i+1}=\vect{X}_i+\vect{a}(\vect{X}_i,\vect{\sigma})(\vect{Z}_{i+1}-\vect{Z}_i)+\frac{\vect{x}_T-\vect{X}_i}{m-i}\label{eq:imputation}
\ee
where $ m-1 $ is the number of imputed points between two
observations. Defining the process $\vect{Z}$ ensures that the
dominating measure is parameter free and,  hence, improves the
performance of the MH sampler. See \citet{Dargatz:2010} for more
details.

We sample $\vect{\sigma}$ according to Algorithm
\ref{alg:diffusionparams}. We use zero-based numbering and $N-1$
observation intervals indexed $0\ldots N-2$. We assume that the
inter-observation times $\Delta$ are all equal and that there are
$m-1$ imputed points per interval, giving a time interval of
$\delta=\Delta/m$. We use the notation $\vect{X}_i=\vect{X}_{t_i}$ and
$\vect{\mu}_i=\vect{\mu}(\vect{X}_{t_i},\vect{A})$. We assume that we
have perfect observations for ease of notation but our method can be
extended to the case of measurement error
(e.g. \cite{Golightly:2008}). The extension to
variable inter-observation times is straight forward. For simplicity, we write
the algorithm for perfect observation of the system so that $\vect{X}_{im},
i=0,\ldots ,N-1$ are fixed. In Algorithm \ref{alg:diffusionparams}, $\phi$
denotes a Gaussian distribution and $q$ the Gaussian proposal density (which
is defined below in Eq. \ref{eq:linearbridgeprop}) and $\Sigma$ a covariance
matrix (given below in Eq. \ref{eq:linearvar}).

\begin{algorithm}
 \caption{Sample parameters entering the diffusion matrix.}
\label{alg:diffusionparams}
\begin{algorithmic}
  \STATE{Draw $\vect{\sigma}^*\sim q(\vect{\sigma}^*|\vect{\sigma})$}
  \STATE{Initialize $\alpha=\log(q(\vect{\sigma}|\vect{\sigma}^*))-\log(q(\vect{\sigma}^*|\vect{\sigma}))+\log(p(\vect{\sigma}^*))-\log(p(\vect{\sigma}))$}
 \FOR{$i=0$ to $N-2$}
 \FOR{$j=0$ to $m-2$}
\STATE{$\vect{Z}_{im+j+1}=\vect{Z}_{im+j}+\vect{a}^{-1}(\vect{X}_{im+j},\vect{\sigma})\left(\vect{X}_{im+j+1}-\vect{X}_{im+j}-\frac{\vect{X}_{im+m}-\vect{X}_{im+j}}{m-j}\right)$} 
\vspace{5mm}
\STATE{$\vect{X}_{im+j+1}^*=\vect{X}_{im+j}^*+\frac{\vect{X}_{im+m}-\vect{X}_{im+j}^*}{m-j}+\vect{a}(\vect{X}_{im+j}^*,\vect{\sigma}^*)(\vect{Z}_{im+j+1}-\vect{Z}_{im+j})$}
\vspace{-2mm}
\STATE{\begin{align}\hspace{-70mm}\alpha=\alpha&+\log(\phi(\vect{X}_{im+j+1}^*;\vect{X}_{im+j}^*+\vect{\mu}^*_{im+j}\delta,\delta\vect{\Sigma}^*_{im+j})+\log|a(\vect{X}_{im+j}^*,\vect{\sigma}^*)|\nn\\
&-\log(\phi(\vect{X}_{im+j+1};\vect{X}_{im+j}+\vect{\mu}_{im+j}\delta,\delta\vect{\Sigma}_{im+j})-\log|a(\vect{X}_{im+j},\vect{\sigma})|\nn\end{align}}
\ENDFOR
\ENDFOR
\STATE{Set $\{\vect{\sigma},\vect{X}\}=\{\vect{\sigma}^*,\vect{X}^*\}$ with probability $\mbox{min}(1,\exp(\alpha))$
else retain $\{\vect{\sigma},\vect{X}\}$}
\end{algorithmic}
\end{algorithm}

To update missing data between observations we use an independence sampler as
in \cite{Roberts:2001} using the proposal process 
\be
d\vect{X}^*=\vect{\xi}(\vect{X}^*,\vect{X}_T)dt+\vect{a}(\vect{X}^*,\vect{\sigma})d\vect{W}^*\,\label{eq:independencesampler2}\,,
\ee
where $\vect{X}_T$ is the next observation, $\vect{X}^*$ the proposed data,
and where $\vect{a}(\vect{X}^*,\vect{\sigma})$ is the same diffusion function
as that in Eq. (\ref{EQ:1}). $\vect{\xi}$ denotes the modified linear bridge
(see below in section 3.1). The proposal process Eq. (\ref{eq:independencesampler2}) will
have a measure
that is absolutely continuous with respect to the target process in
Eq. (\ref{EQ:1}) because of their common diffusion
function. To update all of the missing data, we propose a block at a time from
Eq. (\ref{eq:independencesampler2}) and then accept the proposed block according to the MH
ratio. If the inter-observation interval is large then the acceptance rate may
become very low and so one may sub-sample smaller blocks. 

For some interval $i$ we set $\vect{X}^*_0=\vect{X}_{im}$ and
$\vect{X}^*_m=\vect{X}_{(i+1)m}$ then we propose
$\vect{X}^*_1:\vect{X}^*_{m-1}$ and accept or reject the block using the
MH acceptance probability \be
\alpha=\frac{p_\delta(\vect{X}^*_{m}|\vect{X}^*_{m-1},\vect{A})\prod_{j=0}^{m-2}p_\delta(\vect{X}^*_{j+1}|\vect{X}^*_j,\vect{A})
  q_\delta(\vect{X}_{im+j+1}|\vect{X}_{im+j},\vect{\xi},\vect{\sigma})}{p_\delta(\vect{X}_{(i+1)m}|\vect{X}_{im+m-1},\vect{A})\prod_{j=0}^{m-2}
  p_\delta(\vect{X}_{im+j+1}|\vect{X}_{im+j},\vect{A})q_\delta(\vect{X}^*_{j+1}|\vect{X}^*_j,\vect{\xi},\vect{\sigma})}\,,\ee
where $p_\delta$ is the transition density of the target
\be
d\vect{X}_t=\vect{\mu}(\vect{X}_t,\vect{A})dt+\vect{a}(\vect{X}_t,\vect{\sigma})d\vect{W}_t\,,\,\,\vect{X}_0=\vect{x}_0\,,\,\,t\in[0,T]\label{eq:sdeinference}\ee
under the Euler approximation over the time interval $\delta$ and where $q_\delta$ denotes the
transition density of the proposal. We choose proposal processes so that given
$\vect{X}_j^*$, $\vect{X}^*_{j+1}$ is approximately Gaussian
distributed. However, Eq. (\ref{eq:independencesampler2}) is not a true Gaussian
process because of the state dependent noise term. Details for updating the
missing data are given in Algorithm \ref{alg:updatedata}.

\begin{algorithm}
 \caption{Sample missing data between observations.}
\label{alg:updatedata}
\begin{algorithmic}
 \FOR{$i=0$ to $N-2$}
 \STATE{Set $\vect{X}^*_0=\vect{X}_{im}$}
 \STATE{Set $\alpha=0$}
 \FOR{$j=0$ to $m-2$}
\STATE{$\vect{X}^*_{j+1}\sim q_\delta(\vect{X}^*_{j+1}|\vect{\xi}(\vect{X}^*_j,\vect{X}_{im+m}),\vect{\sigma})$} 
\vspace{-5mm}
\STATE{\begin{align}&\alpha=\alpha +\log(\phi(\vect{X}_{j+1}^*;\vect{X}^*_{j}+\delta\vect{\mu}_j^*,\delta\vect{\Sigma}_j^*)
+\log(q_\delta(\vect{X}_{im+j+1}|\vect{\xi}(\vect{X}_{im+j},\vect{X}_{im+m}),\vect{\sigma}))\nn\\
&-\log(\phi(\vect{X}_{im+j+1};\vect{X}_{im+j}+\delta\vect{\mu}_{im+j},\delta\vect{\Sigma}_{im+j}))-\log(q_\delta(\vect{X}^*_{j+1}|\vect{X}^*_j,\vect{\xi}(\vect{X}^*_j,\vect{X}_{im+m}),\vect{\sigma}))\nn\end{align}}
\vspace{-5mm}
\ENDFOR
\STATE{$\alpha=\alpha+\log(\phi(\vect{X}_{im+m};\vect{X}^*_{m-1}+\delta\vect{\mu}_{m-1}^*,\delta\vect{\Sigma}_{m-1}^*))$}
\STATE{$-\log(\phi(\vect{X}_{im+m};\vect{X}_{im+m-1}+\delta\vect{\mu}_{im+m-1},\delta\vect{\Sigma}_{im+m-1}))$}
\IF{$\exp(\alpha)>\mathcal{U}(0,1)$}
\FOR{$j=0$ to $m-2$}
\STATE{$\vect{X}_{im+j+1}=\vect{X}_{j+1}^*$}
\ENDFOR
\ENDIF
\ENDFOR
\end{algorithmic}
\end{algorithm}
Algorithms \ref{alg:diffusionparams} and \ref{alg:updatedata} are combined
with standard MH updates for the parameters $\vect{A}$
entering into the drift function. First we update the diffusion
parameters using Algorithm 1, then we update all imputed data. After
that the drift parameters will be updated (see section 3.2). In both
algorithms $X^*$ denotes the proposal which will be generated
dependending on the availability of observations. One could use
Random-Walk proposals but in our case of polynomial models it is more
efficient to implement another Gibbs sampling step. Repeatedly
alternating between these three steps will produce MCMC samples that
can be used to estimate the parameters. In practice we increase the
amount of missing data $m$ until we see convergence in the marginal
distributions of the parameters.

\subsection{Sampling of Diffusion Paths}
Because we want to impute missing data we need efficient methods for
simulating diffusion paths from
Eq. (\ref{EQ:0}) that are conditioned upon given start
$\vect{X}_0=\vect{x}_0$ and end $\vect{X}_m=\vect{x}_m$ points. We
consider the total time interval $\tau_m-\tau_0=\Delta$ divided into
$m$ equidistant sub-intervals so that $\tau_{k+1}-\tau_k=\Delta/m=\delta$.

Having $N$ observations, for each of $i=0,1,2,\ldots N-1$, $X_{im}$ is an
observation. Between every pair of observations the diffusion bridge will need
to be simulated. We use an independence sampler with proposal density of the form
$q(\vect{X}^*|\vect{X})=q(\vect{X}^*)$. Here, we consider proposal processes
of the form of Eq. (\ref{eq:independencesampler2}).

We use a Modified Linear Bridge proposal for sampling of
parameters of the drift of equations of the form of Eq. \ref{EQ:1}. For this
purpose we apply Ito's formula to the drift function  of
Eq. \ref{EQ:1}. This gives the approximating process
\begin{equation}
d \vect{Z}_t = (\vect{Q(X)Z}_t + \vect{r(X},t))dt + \vect{\Sigma(X)}d\vect{B}_t.
\end{equation}
with
\begin{align}
&Q_{ij}=\frac{\partial\mu_i(\vect{X}_s)}{\partial x_j}\nn\\
&r_i(t)=\mu_i(\vect{X}_s)-\sum_j\frac{\partial\mu_i}{\partial x_j}\vect{X}_j(s)+\frac{1}{2}\sum_{j,k,l}a_{jl}(\vect{X}_t)a_{kl}(\vect{X}_t)\frac{\partial^2\mu_i}{\partial x_j\partial x_k}(\vect{X}_t)(t-s)\nn\\
&\vect{\Sigma} = \vect{a}(\vect{X}_s)\nn
\end{align}

This is a local linearization of the nonlinear diffusion over a small time
window \cite{Ozaki:1992,Shoji:1998}. 
First we construct bridge distributions for general multivariate linear
diffusions \citep{Barczykern:2013}. If at time $s$ we have
$\vect{X}_s=\vect{d}$ and at time $T$, $\vect{X}_T=\vect{e}$ then the
distribution of $\vect{X}_t$ for $0\leq s<t\leq T$ can be shown to be
Gaussian with mean \be
\vect{\nu}_{\vect{d},\vect{e}}(s,t)=\vect{\Gamma}(t,T)\vect{\Gamma}(s,T)^{-1}\vect{m}_{\vect{d}}^+(s,t)+\vect{\Gamma}(s,T)^T(\vect{\Gamma}(s,T)^T)^{-1}\vect{m}_{\vect{e}}^-(t,T)\,,\label{eq:linearmean}\ee
where \[
\vect{\Gamma}(s,t)=\int_s^te^{(s-u)\vect{Q}}\vect{\Sigma}\vect{\Sigma}^Te^{(t-u)\vect{Q}^T}du\,,\] \[\vect{m}_{\vect{x}}^+(s,t)=\vect{x}+\int_s^te^{(s-u)\vect{Q}}\vect{r}(u)du\,\,\,\,\mbox{and}\,\,\,\,
\vect{m}_{\vect{x}}^-(s,t)=\vect{x}-\int_s^te^{(t-u)\vect{Q}}\vect{r}(u)du\,.\] 
The covariance matrix is given by \be
\vect{\Sigma}(s,t)=\vect{\Gamma}(t,T)\vect{\Gamma}(s,T)^{-1}\vect{\Gamma}(s,t)\,.\label{eq:linearvar}\ee

In general this matrix can be computed as follows: if we diagonalize
$\vect{Q}$ so that $\vect{Q}=\vect{U}\vect{\Lambda} \vect{U}^{-1}$ then
compute the matrix $\vect{A}$ with components 
\be
V_{ij}=\frac{(\vect{U}^{-1}\vect{\Sigma}\vect{\Sigma}^T\vect{U}^{-T})_{ij}}{\Lambda_{ii}+\Lambda_{jj}}\left(e^{(t-s)\Lambda_{jj}}-e^{(s-t)\Lambda_{ii}}\right),\label{eq:matrixA}
\ee
then
\be \vect{\Gamma}(s,t)=\vect{U}\vect{V}\vect{U}^T \,.\label{eq:gammanocommute}\ee

The proposal distribution is given by
\be 
q(\vect{X}_{k+1}|\vect{X}_k,\vect{X}_m,\vect{V})=\phi\left(\vect{X}_{k+1};\vect{\nu}_{\vect{x}_k,\vect{x}_m}(k\delta,(k+1)\delta),\vect{\Sigma}_{\vect{x}_k}(k\delta,(k+1)\delta)\right)\,,\label{eq:linearbridgeprop}
\ee
where $\vect{\nu}_{\vect{x}_j,\vect{x}_m}(j\delta,(j+1)\delta)$ and
$\vect{\Sigma}_{\vect{x}_0}(j\delta,(j+1)\delta)$ are given in
Eqs. (\ref{eq:linearmean}) and (\ref{eq:linearvar}) respectively.

In contrast to a linear bridge sampler here we update at each imputed
point. This means recomputing the matrices $\vect{\Gamma}(s,t)$ at
each point, although $\vect{Q}$ and $\vect{U}$ are only calculated
once.

\subsection{Inference for Drift Parameters\label{sec:drift}}

Now we give details of the computational implementation of the sampling of
parameters in the drift function. Since the drift parameters enter linearly we
can construct a Gibbs sampler where their conditional posterior is
Gaussian. This greatly improves the mixing of the Markov Chain.

\subsubsection{Gibbs Sampler\label{sec:gibbs}}
Consider $N$ observations with time interval $\delta$. We set
$\vect{Y}_t=\vect{X}_{t+1}-\vect{X}_{t}$ and let
$\vect{U}\in\mathbb{R}^{N-1\times P}$ be the design matrix of the data, scaled
by $\delta$. The columns of $\vect{U}$ are indexed in the same way as the
columns of the parameter matrix $\vect{A}$. For example, a two
dimensional system would have $P=10$ and the following design matrix 
\[ U=\delta\left(\begin{array}{ccccccc} 1 & X_{\scriptscriptstyle 1,1} & X_{\scriptscriptstyle 1,2} & X_{\scriptscriptstyle 1,1}^2 
& X_{\scriptscriptstyle 1,1}X_{\scriptscriptstyle 1,2} & X_{\scriptscriptstyle 1,2}^2 & X_{\scriptscriptstyle 1,1}^3\\
\vdots & \vdots & \vdots & \vdots & \vdots & \vdots & \vdots \\
1 & X_{\scriptscriptstyle N-1,1} & X_{\scriptscriptstyle N-1,2} & X_{\scriptscriptstyle N-1,1}^2 & X_{\scriptscriptstyle N-1,1}X_{\scriptscriptstyle N-1,2} 
      & X_{\scriptscriptstyle N-1,2}^2 & X_{\scriptscriptstyle N-1,1}^3\end{array}\right.\]

\[\left.\begin{array}{ccc} X_{\scriptscriptstyle 1,1}^2X_{\scriptscriptstyle 1,2}& X_{\scriptscriptstyle 1,1}X_{\scriptscriptstyle 1,2}^2 & X_{\scriptscriptstyle 1,2}^3 \\
\vdots & \vdots & \vdots \\
X_{\scriptscriptstyle N-1,1}^2X_{\scriptscriptstyle N-1,2} & X_{\scriptscriptstyle N-1,1}X_{\scriptscriptstyle N-1,2}^2 & X_{\scriptscriptstyle N-1,2}^3\end{array}\right)\]

The log likelihood can be written \be
L(\vect{A};\vect{X})=-\frac{1}{2}\sum_{t=1}^{N-1}|\vect{\Sigma}_{Drift~t}|-\frac{1}{2}\sum_{t=1}^{N-1}\sum_{i,j=1}^D\left(\vect{Y}_{ti}-\sum_{k=1}^PU_{tk}A_{ik}\right)\Sigma^{-1}_{Drift~tij}
\left(Y_{tj}-\sum_{k=1}^PU_{tk}A_{jk}\right), \ee
where the instantaneous covariance matrix $\vect{\Sigma}_{Drift~t}$ is computed from
$\Sigma_{Drift~t,j,k}^{1/2}=(d_{j,k}+\sum_{l=1}^De_{l,j,k}X_{t,j})\Delta^{1/2}$.

We have $DP$ parameters to infer in the matrix $\vect{A}$. We use a zero mean
Gaussian prior with covariance matrix $\vect{\Gamma_{Drift}}\in\mathbb{R}^{DP\times
  DP}$. Let $\Lambda\in\mathbb{R}^{DP\times DP}$, be a matrix with
components \be \Lambda_{(i-1)P+j,(k-1)P+l} =
\sum_{t=1}^{N-1}U_{tj}\Sigma^{-1}_{Drift~tik}U_{tl}+\Gamma^{-1}_{Drift~(i-1)P+j,(k-1)P+l} \ee
where $i,k=1\ldots D$ and $j,l=1\ldots P$. Let $e\in\mathbb{R}^{DP}$ with
components \be e_{(i-1)P+j}=\sum_{t,k}U_{t,j}\Sigma^{-1}_{Drift~tik}Y_{tk}. \ee The
posterior mean $\mu_{(i-1)P+j}$ of $A_{i,j}$ is given by the solution of
$\vect{\Lambda}\vect{\mu}=\vect{b}$ and the posterior covariance is
$\mbox{Cov}(A_{i,j},A_{k,l})=\Lambda^{-1}_{(i-1)P+j,(k-1)P+l}$.

\begin{figure}
 \begin{center}
   \epsfig{file=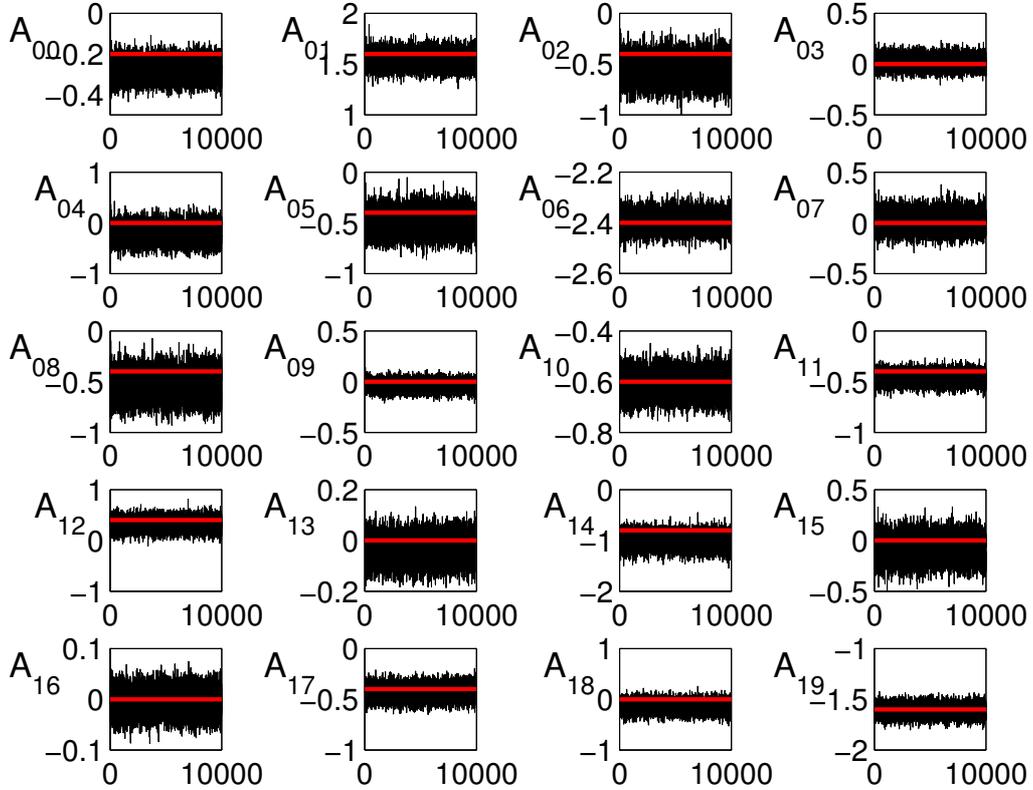,angle=0,scale=0.8}
 \end{center}
\caption{Output of a Gibbs sampler for $20$ drift parameters of a two dimensional
  model from Eq. (\ref{EQ:0}). The observation interval is
  $\delta=10^{-3}$ and $T=10,000$. The true values are shown in red.}
\label{fig:gibbsoutput}
\end{figure}

We applied the above Gibbs sampler to a large data set from a two dimensional
model of the form of Eq. (\ref{EQ:0}) with random values for the
diffusion parameters. We chose a fine observation interval of
$\delta=10^{-3}$ and long observation period
$T=10,000$. Fig. \ref{fig:gibbsoutput} displays the trace plots for
all $20$ parameters (note that the indices are from $0$ rather than
$1$ as in the text). Using this large data set the algorithm is able
to reproduce the true values shown in red.

We performed a further simulation study to test the dependence of the
posterior estimates upon the data set used. We inferred all of the drift
parameters for a simple two dimensional model of the form of
Eq. (\ref{EQ:0}) using data sets of length
$T=\{10,100,1000\}$ and with observation interval
$\Delta=\{0.1,0.01,0.001\}$. Note that the diffusion function is arbitrary in
this model. The results are shown in Table \ref{table:driftinfer}. For each
parameter we estimated the posterior mean and the posterior 10th-90th
percentiles. A cell is colored blue where the true value falls within this
range. Note that if we were using the true likelihood, rather than an
approximation, then we would expect there to be around $80\%$ blue boxes. The
error of the estimates for each data set can be quantified using the quadratic
\textbf{Posterior Expected Loss} (PEL) function 
\be
f(\hat{\pi},\vect{X}_{\mbox{\tiny{obs}}})=\int_{\Theta}(\vect{\theta}^*-\vect{\theta})^2\hat{\pi}(\vect{\theta}|\vect{X}_{\mbox{\tiny{obs}}})d\vect{\theta}\,,
\ee
where $\vect{\theta}$ represents all of the parameters, $\hat{\pi}$ is the
estimated posterior distribution and $\vect{\theta}^*$ is the true value of
the parameter.

\begin{table}
\centering
\scalebox{1.0}{
\renewcommand{\arraystretch}{0.7}\centering\begin{tabular}{l || >{\centering}p{1.0cm} | >{\centering}p{1.0cm} | >{\centering}p{1.0cm} | >{\centering}p{1.0cm} | >{\centering}p{1.0cm} | >{\centering}p{1.0cm} | >{\centering}p{1.0cm} | >{\centering}p{1.0cm} | >{\centering}p{1.0cm}}
 & \multicolumn{3}{c}{$T=10$} & \multicolumn{3}{c}{$T=100$} & \multicolumn{3}{c}{$T=1000$}\tabularnewline

\cline{2-10}
 
 & $\scriptstyle{0.1}$ & $\scriptstyle{0.01}$ & $\scriptstyle{0.001}$ & $\scriptstyle{0.1}$ & $\scriptstyle{0.01}$ & $\scriptstyle{0.001}$ & $\scriptstyle{0.1}$ & $\scriptstyle{0.01}$ & $\scriptstyle{0.001}$ \tabularnewline 
\hline\hline
$A_{00}=0$ &\cellcolor{LightCyan}$\hspace{-3mm}\begin{array}{c}\scriptscriptstyle{2.18}\\ \scriptscriptstyle{(-2.5,6.89)}\end{array}$ &\cellcolor{LightCyan}$\hspace{-3mm}\begin{array}{c}\scriptscriptstyle{-0.44}\\ \scriptscriptstyle{(-4.79,3.74)}\end{array}$ &\cellcolor{LightCyan}$\hspace{-3mm}\begin{array}{c}\scriptscriptstyle{-1.95}\\ \scriptscriptstyle{(-6.07,2.2)}\end{array}$ &\cellcolor{LightCyan}$\hspace{-3mm}\begin{array}{c}\scriptscriptstyle{-0.13}\\ \scriptscriptstyle{(-1.07,0.77)}\end{array}$ &\cellcolor{LightCyan}$\hspace{-3mm}\begin{array}{c}\scriptscriptstyle{0.65}\\ \scriptscriptstyle{(-0.29,1.6)}\end{array}$ &\cellcolor{LightCyan}$\hspace{-3mm}\begin{array}{c}\scriptscriptstyle{0.64}\\ \scriptscriptstyle{(-0.28,1.58)}\end{array}$ &\cellcolor{LightCyan}$\hspace{-3mm}\begin{array}{c}\scriptscriptstyle{0.02}\\ \scriptscriptstyle{(-0.13,0.17)}\end{array}$ &\cellcolor{LightCyan}$\hspace{-3mm}\begin{array}{c}\scriptscriptstyle{0.01}\\ \scriptscriptstyle{(-0.13,0.17)}\end{array}$ &\cellcolor{LightCyan}$\hspace{-3mm}\begin{array}{c}\scriptscriptstyle{0.02}\\ \scriptscriptstyle{(-0.13,0.18)}\end{array}$\tabularnewline \hline
$A_{01}=5$ & $\hspace{-3mm}\begin{array}{c}\scriptscriptstyle{-2.04}\\ \scriptscriptstyle{(-7.23,3.04)}\end{array}$ &\cellcolor{LightCyan}$\hspace{-3mm}\begin{array}{c}\scriptscriptstyle{4.86}\\ \scriptscriptstyle{(0.84,9.04)}\end{array}$ &\cellcolor{LightCyan}$\hspace{-3mm}\begin{array}{c}\scriptscriptstyle{6.21}\\ \scriptscriptstyle{(2.31,10.12)}\end{array}$ & $\hspace{-3mm}\begin{array}{c}\scriptscriptstyle{2.54}\\ \scriptscriptstyle{(1.78,3.28)}\end{array}$ &\cellcolor{LightCyan}$\hspace{-3mm}\begin{array}{c}\scriptscriptstyle{4.73}\\ \scriptscriptstyle{(3.93,5.51)}\end{array}$ &\cellcolor{LightCyan}$\hspace{-3mm}\begin{array}{c}\scriptscriptstyle{5.22}\\ \scriptscriptstyle{(4.43,6.03)}\end{array}$ & $\hspace{-3mm}\begin{array}{c}\scriptscriptstyle{2.84}\\ \scriptscriptstyle{(2.7,2.97)}\end{array}$ & $\hspace{-3mm}\begin{array}{c}\scriptscriptstyle{4.59}\\ \scriptscriptstyle{(4.45,4.73)}\end{array}$ & $\hspace{-3mm}\begin{array}{c}\scriptscriptstyle{4.86}\\ \scriptscriptstyle{(4.73,5)}\end{array}$\tabularnewline \hline
$A_{02}=0$ &\cellcolor{LightCyan}$\hspace{-3mm}\begin{array}{c}\scriptscriptstyle{2.63}\\ \scriptscriptstyle{(-0.16,5.39)}\end{array}$ &\cellcolor{LightCyan}$\hspace{-3mm}\begin{array}{c}\scriptscriptstyle{2.15}\\ \scriptscriptstyle{(-0.17,4.49)}\end{array}$ &\cellcolor{LightCyan}$\hspace{-3mm}\begin{array}{c}\scriptscriptstyle{1.63}\\ \scriptscriptstyle{(-0.66,4)}\end{array}$ &\cellcolor{LightCyan}$\hspace{-3mm}\begin{array}{c}\scriptscriptstyle{0.45}\\ \scriptscriptstyle{(-0.52,1.41)}\end{array}$ &\cellcolor{LightCyan}$\hspace{-3mm}\begin{array}{c}\scriptscriptstyle{0.58}\\ \scriptscriptstyle{(-0.42,1.6)}\end{array}$ &\cellcolor{LightCyan}$\hspace{-3mm}\begin{array}{c}\scriptscriptstyle{0.13}\\ \scriptscriptstyle{(-0.89,1.13)}\end{array}$ &\cellcolor{LightCyan}$\hspace{-3mm}\begin{array}{c}\scriptscriptstyle{-0.08}\\ \scriptscriptstyle{(-0.22,0.05)}\end{array}$ & $\hspace{-3mm}\begin{array}{c}\scriptscriptstyle{-0.14}\\ \scriptscriptstyle{(-0.27,0)}\end{array}$ & $\hspace{-3mm}\begin{array}{c}\scriptscriptstyle{-0.19}\\ \scriptscriptstyle{(-0.32,-0.04)}\end{array}$\tabularnewline \hline
$A_{03}=0$ &\cellcolor{LightCyan}$\hspace{-3mm}\begin{array}{c}\scriptscriptstyle{1.54}\\ \scriptscriptstyle{(-2.75,5.84)}\end{array}$ &\cellcolor{LightCyan}$\hspace{-3mm}\begin{array}{c}\scriptscriptstyle{-0.17}\\ \scriptscriptstyle{(-3.12,2.76)}\end{array}$ &\cellcolor{LightCyan}$\hspace{-3mm}\begin{array}{c}\scriptscriptstyle{0.26}\\ \scriptscriptstyle{(-2.62,3.2)}\end{array}$ &\cellcolor{LightCyan}$\hspace{-3mm}\begin{array}{c}\scriptscriptstyle{0.27}\\ \scriptscriptstyle{(-0.38,0.93)}\end{array}$ &\cellcolor{LightCyan}$\hspace{-3mm}\begin{array}{c}\scriptscriptstyle{0.01}\\ \scriptscriptstyle{(-0.72,0.73)}\end{array}$ &\cellcolor{LightCyan}$\hspace{-3mm}\begin{array}{c}\scriptscriptstyle{-0.17}\\ \scriptscriptstyle{(-0.88,0.54)}\end{array}$ &\cellcolor{LightCyan}$\hspace{-3mm}\begin{array}{c}\scriptscriptstyle{-0.01}\\ \scriptscriptstyle{(-0.08,0.06)}\end{array}$ &\cellcolor{LightCyan}$\hspace{-3mm}\begin{array}{c}\scriptscriptstyle{0}\\ \scriptscriptstyle{(-0.07,0.06)}\end{array}$ &\cellcolor{LightCyan}$\hspace{-3mm}\begin{array}{c}\scriptscriptstyle{0}\\ \scriptscriptstyle{(-0.07,0.06)}\end{array}$\tabularnewline \hline
$A_{04}=0$ &\cellcolor{LightCyan}$\hspace{-3mm}\begin{array}{c}\scriptscriptstyle{-3.51}\\ \scriptscriptstyle{(-8.17,1.13)}\end{array}$ &\cellcolor{LightCyan}$\hspace{-3mm}\begin{array}{c}\scriptscriptstyle{0.61}\\ \scriptscriptstyle{(-3,4.3)}\end{array}$ &\cellcolor{LightCyan}$\hspace{-3mm}\begin{array}{c}\scriptscriptstyle{1.3}\\ \scriptscriptstyle{(-2.32,5.07)}\end{array}$ &\cellcolor{LightCyan}$\hspace{-3mm}\begin{array}{c}\scriptscriptstyle{-0.96}\\ \scriptscriptstyle{(-2.09,0.2)}\end{array}$ &\cellcolor{LightCyan}$\hspace{-3mm}\begin{array}{c}\scriptscriptstyle{-0.58}\\ \scriptscriptstyle{(-1.8,0.68)}\end{array}$ &\cellcolor{LightCyan}$\hspace{-3mm}\begin{array}{c}\scriptscriptstyle{-0.07}\\ \scriptscriptstyle{(-1.3,1.16)}\end{array}$ &\cellcolor{LightCyan}$\hspace{-3mm}\begin{array}{c}\scriptscriptstyle{-0.01}\\ \scriptscriptstyle{(-0.04,0.01)}\end{array}$ &\cellcolor{LightCyan}$\hspace{-3mm}\begin{array}{c}\scriptscriptstyle{-0.02}\\ \scriptscriptstyle{(-0.04,0.01)}\end{array}$ &\cellcolor{LightCyan}$\hspace{-3mm}\begin{array}{c}\scriptscriptstyle{-0.02}\\ \scriptscriptstyle{(-0.05,0.01)}\end{array}$\tabularnewline \hline
$A_{05}=0$ &\cellcolor{LightCyan}$\hspace{-3mm}\begin{array}{c}\scriptscriptstyle{0.32}\\ \scriptscriptstyle{(-2.19,2.81)}\end{array}$ &\cellcolor{LightCyan}$\hspace{-3mm}\begin{array}{c}\scriptscriptstyle{0}\\ \scriptscriptstyle{(-2.31,2.31)}\end{array}$ &\cellcolor{LightCyan}$\hspace{-3mm}\begin{array}{c}\scriptscriptstyle{0.19}\\ \scriptscriptstyle{(-2.11,2.49)}\end{array}$ &\cellcolor{LightCyan}$\hspace{-3mm}\begin{array}{c}\scriptscriptstyle{0.66}\\ \scriptscriptstyle{(-0.11,1.43)}\end{array}$ &\cellcolor{LightCyan}$\hspace{-3mm}\begin{array}{c}\scriptscriptstyle{0.02}\\ \scriptscriptstyle{(-0.79,0.84)}\end{array}$ &\cellcolor{LightCyan}$\hspace{-3mm}\begin{array}{c}\scriptscriptstyle{-0.27}\\ \scriptscriptstyle{(-1.05,0.55)}\end{array}$ &\cellcolor{LightCyan}$\hspace{-3mm}\begin{array}{c}\scriptscriptstyle{0}\\ \scriptscriptstyle{(-0.06,0.07)}\end{array}$ &\cellcolor{LightCyan}$\hspace{-3mm}\begin{array}{c}\scriptscriptstyle{0.01}\\ \scriptscriptstyle{(-0.06,0.07)}\end{array}$ &\cellcolor{LightCyan}$\hspace{-3mm}\begin{array}{c}\scriptscriptstyle{0}\\ \scriptscriptstyle{(-0.07,0.07)}\end{array}$\tabularnewline \hline
$A_{06}=-3$ & $\hspace{-3mm}\begin{array}{c}\scriptscriptstyle{-0.88}\\ \scriptscriptstyle{(-2.96,1.24)}\end{array}$ &\cellcolor{LightCyan}$\hspace{-3mm}\begin{array}{c}\scriptscriptstyle{-2.58}\\ \scriptscriptstyle{(-4.3,-0.87)}\end{array}$ &\cellcolor{LightCyan}$\hspace{-3mm}\begin{array}{c}\scriptscriptstyle{-3.17}\\ \scriptscriptstyle{(-4.83,-1.49)}\end{array}$ & $\hspace{-3mm}\begin{array}{c}\scriptscriptstyle{-1.8}\\ \scriptscriptstyle{(-2,-1.6)}\end{array}$ &\cellcolor{LightCyan}$\hspace{-3mm}\begin{array}{c}\scriptscriptstyle{-2.92}\\ \scriptscriptstyle{(-3.12,-2.71)}\end{array}$ &\cellcolor{LightCyan}$\hspace{-3mm}\begin{array}{c}\scriptscriptstyle{-3.05}\\ \scriptscriptstyle{(-3.25,-2.85)}\end{array}$ & $\hspace{-3mm}\begin{array}{c}\scriptscriptstyle{-1.71}\\ \scriptscriptstyle{(-1.77,-1.65)}\end{array}$ & $\hspace{-3mm}\begin{array}{c}\scriptscriptstyle{-2.75}\\ \scriptscriptstyle{(-2.81,-2.69)}\end{array}$ & $\hspace{-3mm}\begin{array}{c}\scriptscriptstyle{-2.91}\\ \scriptscriptstyle{(-2.97,-2.85)}\end{array}$\tabularnewline \hline
$A_{07}=0$ &\cellcolor{LightCyan}$\hspace{-3mm}\begin{array}{c}\scriptscriptstyle{0.58}\\ \scriptscriptstyle{(-1.87,3.06)}\end{array}$ &\cellcolor{LightCyan}$\hspace{-3mm}\begin{array}{c}\scriptscriptstyle{-1.32}\\ \scriptscriptstyle{(-3.32,0.7)}\end{array}$ &\cellcolor{LightCyan}$\hspace{-3mm}\begin{array}{c}\scriptscriptstyle{-1.48}\\ \scriptscriptstyle{(-3.54,0.52)}\end{array}$ &\cellcolor{LightCyan}$\hspace{-3mm}\begin{array}{c}\scriptscriptstyle{0.05}\\ \scriptscriptstyle{(-0.49,0.58)}\end{array}$ &\cellcolor{LightCyan}$\hspace{-3mm}\begin{array}{c}\scriptscriptstyle{-0.07}\\ \scriptscriptstyle{(-0.66,0.51)}\end{array}$ &\cellcolor{LightCyan}$\hspace{-3mm}\begin{array}{c}\scriptscriptstyle{-0.22}\\ \scriptscriptstyle{(-0.79,0.37)}\end{array}$ &\cellcolor{LightCyan}$\hspace{-3mm}\begin{array}{c}\scriptscriptstyle{-0.02}\\ \scriptscriptstyle{(-0.07,0.04)}\end{array}$ &\cellcolor{LightCyan}$\hspace{-3mm}\begin{array}{c}\scriptscriptstyle{-0.01}\\ \scriptscriptstyle{(-0.06,0.05)}\end{array}$ &\cellcolor{LightCyan}$\hspace{-3mm}\begin{array}{c}\scriptscriptstyle{0}\\ \scriptscriptstyle{(-0.05,0.06)}\end{array}$\tabularnewline \hline
$A_{08}=0$ &\cellcolor{LightCyan}$\hspace{-3mm}\begin{array}{c}\scriptscriptstyle{-0.83}\\ \scriptscriptstyle{(-3,1.31)}\end{array}$ &\cellcolor{LightCyan}$\hspace{-3mm}\begin{array}{c}\scriptscriptstyle{-0.28}\\ \scriptscriptstyle{(-2.31,1.76)}\end{array}$ &\cellcolor{LightCyan}$\hspace{-3mm}\begin{array}{c}\scriptscriptstyle{-0.18}\\ \scriptscriptstyle{(-2.21,1.86)}\end{array}$ &\cellcolor{LightCyan}$\hspace{-3mm}\begin{array}{c}\scriptscriptstyle{-0.48}\\ \scriptscriptstyle{(-1.03,0.07)}\end{array}$ &\cellcolor{LightCyan}$\hspace{-3mm}\begin{array}{c}\scriptscriptstyle{-0.37}\\ \scriptscriptstyle{(-0.96,0.23)}\end{array}$ &\cellcolor{LightCyan}$\hspace{-3mm}\begin{array}{c}\scriptscriptstyle{-0.16}\\ \scriptscriptstyle{(-0.75,0.42)}\end{array}$ &\cellcolor{LightCyan}$\hspace{-3mm}\begin{array}{c}\scriptscriptstyle{-0.02}\\ \scriptscriptstyle{(-0.08,0.03)}\end{array}$ &\cellcolor{LightCyan}$\hspace{-3mm}\begin{array}{c}\scriptscriptstyle{-0.01}\\ \scriptscriptstyle{(-0.06,0.04)}\end{array}$ &\cellcolor{LightCyan}$\hspace{-3mm}\begin{array}{c}\scriptscriptstyle{-0.01}\\ \scriptscriptstyle{(-0.07,0.04)}\end{array}$\tabularnewline \hline
$A_{09}=0$ &\cellcolor{LightCyan}$\hspace{-3mm}\begin{array}{c}\scriptscriptstyle{0.71}\\ \scriptscriptstyle{(-0.11,1.53)}\end{array}$ &\cellcolor{LightCyan}$\hspace{-3mm}\begin{array}{c}\scriptscriptstyle{-0.4}\\ \scriptscriptstyle{(-1.16,0.38)}\end{array}$ &\cellcolor{LightCyan}$\hspace{-3mm}\begin{array}{c}\scriptscriptstyle{-0.49}\\ \scriptscriptstyle{(-1.26,0.27)}\end{array}$ &\cellcolor{LightCyan}$\hspace{-3mm}\begin{array}{c}\scriptscriptstyle{0.33}\\ \scriptscriptstyle{(-0.02,0.67)}\end{array}$ &\cellcolor{LightCyan}$\hspace{-3mm}\begin{array}{c}\scriptscriptstyle{0.04}\\ \scriptscriptstyle{(-0.3,0.39)}\end{array}$ &\cellcolor{LightCyan}$\hspace{-3mm}\begin{array}{c}\scriptscriptstyle{0.09}\\ \scriptscriptstyle{(-0.25,0.44)}\end{array}$ & $\hspace{-3mm}\begin{array}{c}\scriptscriptstyle{0.08}\\ \scriptscriptstyle{(0.02,0.13)}\end{array}$ & $\hspace{-3mm}\begin{array}{c}\scriptscriptstyle{0.1}\\ \scriptscriptstyle{(0.04,0.16)}\end{array}$ & $\hspace{-3mm}\begin{array}{c}\scriptscriptstyle{0.12}\\ \scriptscriptstyle{(0.05,0.17)}\end{array}$\tabularnewline \hline
$A_{10}=0$ &\cellcolor{LightCyan}$\hspace{-3mm}\begin{array}{c}\scriptscriptstyle{-0.38}\\ \scriptscriptstyle{(-5.19,4.24)}\end{array}$ &\cellcolor{LightCyan}$\hspace{-3mm}\begin{array}{c}\scriptscriptstyle{-2.21}\\ \scriptscriptstyle{(-6.44,2.03)}\end{array}$ &\cellcolor{LightCyan}$\hspace{-3mm}\begin{array}{c}\scriptscriptstyle{-1.17}\\ \scriptscriptstyle{(-5.21,2.9)}\end{array}$ &\cellcolor{LightCyan}$\hspace{-3mm}\begin{array}{c}\scriptscriptstyle{-0.31}\\ \scriptscriptstyle{(-1.22,0.61)}\end{array}$ &\cellcolor{LightCyan}$\hspace{-3mm}\begin{array}{c}\scriptscriptstyle{-0.18}\\ \scriptscriptstyle{(-1.1,0.75)}\end{array}$ &\cellcolor{LightCyan}$\hspace{-3mm}\begin{array}{c}\scriptscriptstyle{0.06}\\ \scriptscriptstyle{(-0.89,0.99)}\end{array}$ &\cellcolor{LightCyan}$\hspace{-3mm}\begin{array}{c}\scriptscriptstyle{0.02}\\ \scriptscriptstyle{(-0.13,0.17)}\end{array}$ &\cellcolor{LightCyan}$\hspace{-3mm}\begin{array}{c}\scriptscriptstyle{0.05}\\ \scriptscriptstyle{(-0.1,0.2)}\end{array}$ &\cellcolor{LightCyan}$\hspace{-3mm}\begin{array}{c}\scriptscriptstyle{0.05}\\ \scriptscriptstyle{(-0.1,0.2)}\end{array}$\tabularnewline \hline
$A_{11}=0$ &\cellcolor{LightCyan}$\hspace{-3mm}\begin{array}{c}\scriptscriptstyle{-2.93}\\ \scriptscriptstyle{(-8.11,2.15)}\end{array}$ &\cellcolor{LightCyan}$\hspace{-3mm}\begin{array}{c}\scriptscriptstyle{2.42}\\ \scriptscriptstyle{(-1.64,6.48)}\end{array}$ &\cellcolor{LightCyan}$\hspace{-3mm}\begin{array}{c}\scriptscriptstyle{1.67}\\ \scriptscriptstyle{(-2.29,5.64)}\end{array}$ & $\hspace{-3mm}\begin{array}{c}\scriptscriptstyle{0.88}\\ \scriptscriptstyle{(0.1,1.64)}\end{array}$ &\cellcolor{LightCyan}$\hspace{-3mm}\begin{array}{c}\scriptscriptstyle{0.39}\\ \scriptscriptstyle{(-0.4,1.19)}\end{array}$ &\cellcolor{LightCyan}$\hspace{-3mm}\begin{array}{c}\scriptscriptstyle{0.33}\\ \scriptscriptstyle{(-0.48,1.12)}\end{array}$ &\cellcolor{LightCyan}$\hspace{-3mm}\begin{array}{c}\scriptscriptstyle{-0.06}\\ \scriptscriptstyle{(-0.2,0.08)}\end{array}$ &\cellcolor{LightCyan}$\hspace{-3mm}\begin{array}{c}\scriptscriptstyle{0.11}\\ \scriptscriptstyle{(-0.03,0.25)}\end{array}$ &\cellcolor{LightCyan}$\hspace{-3mm}\begin{array}{c}\scriptscriptstyle{0.11}\\ \scriptscriptstyle{(-0.04,0.24)}\end{array}$\tabularnewline \hline
$A_{12}=5$ &\cellcolor{LightCyan}$\hspace{-3mm}\begin{array}{c}\scriptscriptstyle{5.96}\\ \scriptscriptstyle{(3.22,8.68)}\end{array}$ &\cellcolor{LightCyan}$\hspace{-3mm}\begin{array}{c}\scriptscriptstyle{4.2}\\ \scriptscriptstyle{(1.89,6.63)}\end{array}$ &\cellcolor{LightCyan}$\hspace{-3mm}\begin{array}{c}\scriptscriptstyle{4.72}\\ \scriptscriptstyle{(2.33,7.09)}\end{array}$ & $\hspace{-3mm}\begin{array}{c}\scriptscriptstyle{1.68}\\ \scriptscriptstyle{(0.72,2.65)}\end{array}$ &\cellcolor{LightCyan}$\hspace{-3mm}\begin{array}{c}\scriptscriptstyle{4.56}\\ \scriptscriptstyle{(3.52,5.57)}\end{array}$ &\cellcolor{LightCyan}$\hspace{-3mm}\begin{array}{c}\scriptscriptstyle{5.09}\\ \scriptscriptstyle{(4.08,6.12)}\end{array}$ & $\hspace{-3mm}\begin{array}{c}\scriptscriptstyle{2.88}\\ \scriptscriptstyle{(2.74,3.02)}\end{array}$ & $\hspace{-3mm}\begin{array}{c}\scriptscriptstyle{4.7}\\ \scriptscriptstyle{(4.56,4.84)}\end{array}$ &\cellcolor{LightCyan}$\hspace{-3mm}\begin{array}{c}\scriptscriptstyle{4.98}\\ \scriptscriptstyle{(4.84,5.11)}\end{array}$\tabularnewline \hline
$A_{13}=0$ &\cellcolor{LightCyan}$\hspace{-3mm}\begin{array}{c}\scriptscriptstyle{2.68}\\ \scriptscriptstyle{(-1.67,6.93)}\end{array}$ &\cellcolor{LightCyan}$\hspace{-3mm}\begin{array}{c}\scriptscriptstyle{-1.41}\\ \scriptscriptstyle{(-4.36,1.57)}\end{array}$ &\cellcolor{LightCyan}$\hspace{-3mm}\begin{array}{c}\scriptscriptstyle{-1.88}\\ \scriptscriptstyle{(-4.74,1.02)}\end{array}$ & $\hspace{-3mm}\begin{array}{c}\scriptscriptstyle{-1.05}\\ \scriptscriptstyle{(-1.72,-0.38)}\end{array}$ &\cellcolor{LightCyan}$\hspace{-3mm}\begin{array}{c}\scriptscriptstyle{-0.61}\\ \scriptscriptstyle{(-1.33,0.1)}\end{array}$ & $\hspace{-3mm}\begin{array}{c}\scriptscriptstyle{-0.74}\\ \scriptscriptstyle{(-1.46,-0.02)}\end{array}$ &\cellcolor{LightCyan}$\hspace{-3mm}\begin{array}{c}\scriptscriptstyle{-0.01}\\ \scriptscriptstyle{(-0.08,0.05)}\end{array}$ &\cellcolor{LightCyan}$\hspace{-3mm}\begin{array}{c}\scriptscriptstyle{-0.03}\\ \scriptscriptstyle{(-0.1,0.03)}\end{array}$ &\cellcolor{LightCyan}$\hspace{-3mm}\begin{array}{c}\scriptscriptstyle{-0.03}\\ \scriptscriptstyle{(-0.09,0.04)}\end{array}$\tabularnewline \hline
$A_{14}=0$ &\cellcolor{LightCyan}$\hspace{-3mm}\begin{array}{c}\scriptscriptstyle{-1.13}\\ \scriptscriptstyle{(-5.84,3.6)}\end{array}$ & $\hspace{-3mm}\begin{array}{c}\scriptscriptstyle{4.67}\\ \scriptscriptstyle{(1.01,8.45)}\end{array}$ & $\hspace{-3mm}\begin{array}{c}\scriptscriptstyle{4.23}\\ \scriptscriptstyle{(0.47,7.92)}\end{array}$ & $\hspace{-3mm}\begin{array}{c}\scriptscriptstyle{1.39}\\ \scriptscriptstyle{(0.24,2.55)}\end{array}$ &\cellcolor{LightCyan}$\hspace{-3mm}\begin{array}{c}\scriptscriptstyle{0.78}\\ \scriptscriptstyle{(-0.44,2.03)}\end{array}$ &\cellcolor{LightCyan}$\hspace{-3mm}\begin{array}{c}\scriptscriptstyle{0.91}\\ \scriptscriptstyle{(-0.33,2.14)}\end{array}$ &\cellcolor{LightCyan}$\hspace{-3mm}\begin{array}{c}\scriptscriptstyle{0.02}\\ \scriptscriptstyle{(-0.01,0.05)}\end{array}$ & $\hspace{-3mm}\begin{array}{c}\scriptscriptstyle{0.03}\\ \scriptscriptstyle{(0,0.05)}\end{array}$ & $\hspace{-3mm}\begin{array}{c}\scriptscriptstyle{0.03}\\ \scriptscriptstyle{(0,0.05)}\end{array}$\tabularnewline \hline
$A_{15}=0$ &\cellcolor{LightCyan}$\hspace{-3mm}\begin{array}{c}\scriptscriptstyle{0.42}\\ \scriptscriptstyle{(-2.07,2.94)}\end{array}$ &\cellcolor{LightCyan}$\hspace{-3mm}\begin{array}{c}\scriptscriptstyle{-0.6}\\ \scriptscriptstyle{(-2.88,1.77)}\end{array}$ &\cellcolor{LightCyan}$\hspace{-3mm}\begin{array}{c}\scriptscriptstyle{-0.84}\\ \scriptscriptstyle{(-3.13,1.45)}\end{array}$ &\cellcolor{LightCyan}$\hspace{-3mm}\begin{array}{c}\scriptscriptstyle{-0.22}\\ \scriptscriptstyle{(-1.01,0.56)}\end{array}$ &\cellcolor{LightCyan}$\hspace{-3mm}\begin{array}{c}\scriptscriptstyle{-0.11}\\ \scriptscriptstyle{(-0.9,0.7)}\end{array}$ &\cellcolor{LightCyan}$\hspace{-3mm}\begin{array}{c}\scriptscriptstyle{-0.24}\\ \scriptscriptstyle{(-1.06,0.55)}\end{array}$ &\cellcolor{LightCyan}$\hspace{-3mm}\begin{array}{c}\scriptscriptstyle{0}\\ \scriptscriptstyle{(-0.07,0.06)}\end{array}$ &\cellcolor{LightCyan}$\hspace{-3mm}\begin{array}{c}\scriptscriptstyle{-0.01}\\ \scriptscriptstyle{(-0.08,0.05)}\end{array}$ &\cellcolor{LightCyan}$\hspace{-3mm}\begin{array}{c}\scriptscriptstyle{-0.02}\\ \scriptscriptstyle{(-0.09,0.04)}\end{array}$\tabularnewline \hline
$A_{16}=0$ &\cellcolor{LightCyan}$\hspace{-3mm}\begin{array}{c}\scriptscriptstyle{-0.82}\\ \scriptscriptstyle{(-2.94,1.29)}\end{array}$ &\cellcolor{LightCyan}$\hspace{-3mm}\begin{array}{c}\scriptscriptstyle{0.12}\\ \scriptscriptstyle{(-1.61,1.83)}\end{array}$ &\cellcolor{LightCyan}$\hspace{-3mm}\begin{array}{c}\scriptscriptstyle{0.39}\\ \scriptscriptstyle{(-1.26,2)}\end{array}$ &\cellcolor{LightCyan}$\hspace{-3mm}\begin{array}{c}\scriptscriptstyle{-0.08}\\ \scriptscriptstyle{(-0.28,0.12)}\end{array}$ &\cellcolor{LightCyan}$\hspace{-3mm}\begin{array}{c}\scriptscriptstyle{-0.04}\\ \scriptscriptstyle{(-0.24,0.17)}\end{array}$ &\cellcolor{LightCyan}$\hspace{-3mm}\begin{array}{c}\scriptscriptstyle{-0.02}\\ \scriptscriptstyle{(-0.22,0.19)}\end{array}$ &\cellcolor{LightCyan}$\hspace{-3mm}\begin{array}{c}\scriptscriptstyle{-0.03}\\ \scriptscriptstyle{(-0.09,0.03)}\end{array}$ & $\hspace{-3mm}\begin{array}{c}\scriptscriptstyle{-0.08}\\ \scriptscriptstyle{(-0.14,-0.02)}\end{array}$ & $\hspace{-3mm}\begin{array}{c}\scriptscriptstyle{-0.09}\\ \scriptscriptstyle{(-0.15,-0.03)}\end{array}$\tabularnewline \hline
$A_{17}=0$ &\cellcolor{LightCyan}$\hspace{-3mm}\begin{array}{c}\scriptscriptstyle{-0.46}\\ \scriptscriptstyle{(-2.95,2.04)}\end{array}$ & $\hspace{-3mm}\begin{array}{c}\scriptscriptstyle{-3.21}\\ \scriptscriptstyle{(-5.28,-1.22)}\end{array}$ & $\hspace{-3mm}\begin{array}{c}\scriptscriptstyle{-2.97}\\ \scriptscriptstyle{(-4.99,-0.93)}\end{array}$ & $\hspace{-3mm}\begin{array}{c}\scriptscriptstyle{-0.71}\\ \scriptscriptstyle{(-1.25,-0.17)}\end{array}$ & $\hspace{-3mm}\begin{array}{c}\scriptscriptstyle{-0.66}\\ \scriptscriptstyle{(-1.24,-0.08)}\end{array}$ & $\hspace{-3mm}\begin{array}{c}\scriptscriptstyle{-0.78}\\ \scriptscriptstyle{(-1.35,-0.19)}\end{array}$ &\cellcolor{LightCyan}$\hspace{-3mm}\begin{array}{c}\scriptscriptstyle{0.01}\\ \scriptscriptstyle{(-0.05,0.06)}\end{array}$ &\cellcolor{LightCyan}$\hspace{-3mm}\begin{array}{c}\scriptscriptstyle{-0.01}\\ \scriptscriptstyle{(-0.06,0.04)}\end{array}$ &\cellcolor{LightCyan}$\hspace{-3mm}\begin{array}{c}\scriptscriptstyle{0}\\ \scriptscriptstyle{(-0.05,0.05)}\end{array}$\tabularnewline \hline
$A_{18}=0$ &\cellcolor{LightCyan}$\hspace{-3mm}\begin{array}{c}\scriptscriptstyle{0.66}\\ \scriptscriptstyle{(-1.5,2.86)}\end{array}$ &\cellcolor{LightCyan}$\hspace{-3mm}\begin{array}{c}\scriptscriptstyle{1.33}\\ \scriptscriptstyle{(-0.71,3.35)}\end{array}$ &\cellcolor{LightCyan}$\hspace{-3mm}\begin{array}{c}\scriptscriptstyle{1.63}\\ \scriptscriptstyle{(-0.41,3.6)}\end{array}$ & $\hspace{-3mm}\begin{array}{c}\scriptscriptstyle{0.65}\\ \scriptscriptstyle{(0.11,1.22)}\end{array}$ &\cellcolor{LightCyan}$\hspace{-3mm}\begin{array}{c}\scriptscriptstyle{0.45}\\ \scriptscriptstyle{(-0.13,1.04)}\end{array}$ &\cellcolor{LightCyan}$\hspace{-3mm}\begin{array}{c}\scriptscriptstyle{0.58}\\ \scriptscriptstyle{(-0.01,1.18)}\end{array}$ & $\hspace{-3mm}\begin{array}{c}\scriptscriptstyle{0.06}\\ \scriptscriptstyle{(0.01,0.12)}\end{array}$ &\cellcolor{LightCyan}$\hspace{-3mm}\begin{array}{c}\scriptscriptstyle{0.02}\\ \scriptscriptstyle{(-0.03,0.08)}\end{array}$ &\cellcolor{LightCyan}$\hspace{-3mm}\begin{array}{c}\scriptscriptstyle{0.03}\\ \scriptscriptstyle{(-0.02,0.09)}\end{array}$\tabularnewline \hline
$A_{19}=-3$ &\cellcolor{LightCyan}$\hspace{-3mm}\begin{array}{c}\scriptscriptstyle{-2.61}\\ \scriptscriptstyle{(-3.44,-1.78)}\end{array}$ &\cellcolor{LightCyan}$\hspace{-3mm}\begin{array}{c}\scriptscriptstyle{-3.23}\\ \scriptscriptstyle{(-4.02,-2.45)}\end{array}$ &\cellcolor{LightCyan}$\hspace{-3mm}\begin{array}{c}\scriptscriptstyle{-3.47}\\ \scriptscriptstyle{(-4.24,-2.69)}\end{array}$ & $\hspace{-3mm}\begin{array}{c}\scriptscriptstyle{-1.46}\\ \scriptscriptstyle{(-1.8,-1.12)}\end{array}$ &\cellcolor{LightCyan}$\hspace{-3mm}\begin{array}{c}\scriptscriptstyle{-2.73}\\ \scriptscriptstyle{(-3.07,-2.38)}\end{array}$ &\cellcolor{LightCyan}$\hspace{-3mm}\begin{array}{c}\scriptscriptstyle{-3.04}\\ \scriptscriptstyle{(-3.39,-2.7)}\end{array}$ & $\hspace{-3mm}\begin{array}{c}\scriptscriptstyle{-1.75}\\ \scriptscriptstyle{(-1.81,-1.69)}\end{array}$ & $\hspace{-3mm}\begin{array}{c}\scriptscriptstyle{-2.81}\\ \scriptscriptstyle{(-2.87,-2.75)}\end{array}$ &\cellcolor{LightCyan}$\hspace{-3mm}\begin{array}{c}\scriptscriptstyle{-2.98}\\ \scriptscriptstyle{(-3.04,-2.92)}\end{array}$\tabularnewline \hline
 &\cellcolor{Violet} 8.48 & \cellcolor{Violet} 5.25 & \cellcolor{Violet} 4.96 & \cellcolor{Violet} 1.19 & \cellcolor{Violet} 0.37 & \cellcolor{Violet} 0.36 & \cellcolor{Violet} 0.43 & \cellcolor{Violet} 0.02 & \cellcolor{Violet} 0.01\tabularnewline 
\end{tabular}

}
\caption{Drift parameter estimates for a two dimensional cubic model with
  arbitrary diffusion function. On the left is the true value of the
  parameter. The length of the data set used for the inference is labeled as
  $T$ and the observation interval is $\Delta=\{0.1,0.01,0.001\}$. In each
  cell the parameter is estimated from the posterior mean and in brackets is
  shown the 10th-90th percentiles of the posterior. The blue coloring is where
  the true value falls in this range. The bottom of the table shows the
  Posterior Expected Loss in each case.}
\label{table:driftinfer}
\end{table}

We performed a test with both the Gibbs sampler and data
imputation. In Tab. \ref{table:driftinfer2} the data is observed at
interval $\Delta=0.1$. The smaller intervals $\Delta=\{0.01,0.001\}$ are
obtained by imputing data with $m=\{10,100\}$ respectively. The table shows
that imputing data approximately doubles the Posterior Expected Loss. As
expected the confidence intervals are broader but with more imputed data the
algorithm can recover the true values. This shows that our data
imputing strategy successfully improves the parameter estimates.

\begin{table}
\scalebox{1.0}{
\renewcommand{\arraystretch}{0.7}\centering\begin{tabular}{l || >{\centering}p{1.0cm} | >{\centering}p{1.0cm} | >{\centering}p{1.0cm} | >{\centering}p{1.0cm} | >{\centering}p{1.0cm} | >{\centering}p{1.0cm} | >{\centering}p{1.0cm} | >{\centering}p{1.0cm} | >{\centering}p{1.0cm}}
 & \multicolumn{3}{c}{$T=10$} & \multicolumn{3}{c}{$T=100$} & \multicolumn{3}{c}{$T=1000$}\tabularnewline

\cline{2-10}
 
 & $\scriptstyle{0.1}$ & $\scriptstyle{0.01}$ & $\scriptstyle{0.001}$ & $\scriptstyle{0.1}$ & $\scriptstyle{0.01}$ & $\scriptstyle{0.001}$ & $\scriptstyle{0.1}$ & $\scriptstyle{0.01}$ & $\scriptstyle{0.001}$ \tabularnewline 
\hline\hline
$A_{00}=0$ &\cellcolor{LightCyan}$\hspace{-3mm}\begin{array}{c}\scriptscriptstyle{2.18}\\ \scriptscriptstyle{(-2.5,6.89)}\end{array}$ &\cellcolor{LightCyan}$\hspace{-3mm}\begin{array}{c}\scriptscriptstyle{-2.46}\\ \scriptscriptstyle{(-8.46,3.52)}\end{array}$ &\cellcolor{LightCyan}$\hspace{-3mm}\begin{array}{c}\scriptscriptstyle{-2.59}\\ \scriptscriptstyle{(-8.35,3.24)}\end{array}$ &\cellcolor{LightCyan}$\hspace{-3mm}\begin{array}{c}\scriptscriptstyle{-0.13}\\ \scriptscriptstyle{(-1.07,0.77)}\end{array}$ &\cellcolor{LightCyan}$\hspace{-3mm}\begin{array}{c}\scriptscriptstyle{-0.68}\\ \scriptscriptstyle{(-1.84,0.5)}\end{array}$ &\cellcolor{LightCyan}$\hspace{-3mm}\begin{array}{c}\scriptscriptstyle{-0.69}\\ \scriptscriptstyle{(-1.85,0.49)}\end{array}$ &\cellcolor{LightCyan}$\hspace{-3mm}\begin{array}{c}\scriptscriptstyle{0.02}\\ \scriptscriptstyle{(-0.13,0.17)}\end{array}$ &\cellcolor{LightCyan}$\hspace{-3mm}\begin{array}{c}\scriptscriptstyle{-0.08}\\ \scriptscriptstyle{(-0.27,0.09)}\end{array}$ &\cellcolor{LightCyan}$\hspace{-3mm}\begin{array}{c}\scriptscriptstyle{-0.08}\\ \scriptscriptstyle{(-0.26,0.09)}\end{array}$\tabularnewline \hline
$A_{01}=5$ & $\hspace{-3mm}\begin{array}{c}\scriptscriptstyle{-2.04}\\ \scriptscriptstyle{(-7.23,3.04)}\end{array}$ &\cellcolor{LightCyan}$\hspace{-3mm}\begin{array}{c}\scriptscriptstyle{4.99}\\ \scriptscriptstyle{(-0.7,10.61)}\end{array}$ &\cellcolor{LightCyan}$\hspace{-3mm}\begin{array}{c}\scriptscriptstyle{4.84}\\ \scriptscriptstyle{(-0.49,10.22)}\end{array}$ & $\hspace{-3mm}\begin{array}{c}\scriptscriptstyle{2.54}\\ \scriptscriptstyle{(1.78,3.28)}\end{array}$ &\cellcolor{LightCyan}$\hspace{-3mm}\begin{array}{c}\scriptscriptstyle{4.89}\\ \scriptscriptstyle{(3.83,5.97)}\end{array}$ &\cellcolor{LightCyan}$\hspace{-3mm}\begin{array}{c}\scriptscriptstyle{4.76}\\ \scriptscriptstyle{(3.74,5.79)}\end{array}$ & $\hspace{-3mm}\begin{array}{c}\scriptscriptstyle{2.84}\\ \scriptscriptstyle{(2.7,2.97)}\end{array}$ &\cellcolor{LightCyan}$\hspace{-3mm}\begin{array}{c}\scriptscriptstyle{5.04}\\ \scriptscriptstyle{(4.87,5.21)}\end{array}$ &\cellcolor{LightCyan}$\hspace{-3mm}\begin{array}{c}\scriptscriptstyle{4.88}\\ \scriptscriptstyle{(4.71,5.03)}\end{array}$\tabularnewline \hline
$A_{02}=0$ &\cellcolor{LightCyan}$\hspace{-3mm}\begin{array}{c}\scriptscriptstyle{2.63}\\ \scriptscriptstyle{(-0.16,5.39)}\end{array}$ &\cellcolor{LightCyan}$\hspace{-3mm}\begin{array}{c}\scriptscriptstyle{0.61}\\ \scriptscriptstyle{(-2.94,4.05)}\end{array}$ &\cellcolor{LightCyan}$\hspace{-3mm}\begin{array}{c}\scriptscriptstyle{0.39}\\ \scriptscriptstyle{(-2.95,3.75)}\end{array}$ &\cellcolor{LightCyan}$\hspace{-3mm}\begin{array}{c}\scriptscriptstyle{0.45}\\ \scriptscriptstyle{(-0.52,1.41)}\end{array}$ &\cellcolor{LightCyan}$\hspace{-3mm}\begin{array}{c}\scriptscriptstyle{-0.06}\\ \scriptscriptstyle{(-1.42,1.3)}\end{array}$ &\cellcolor{LightCyan}$\hspace{-3mm}\begin{array}{c}\scriptscriptstyle{-0.2}\\ \scriptscriptstyle{(-1.49,1.13)}\end{array}$ &\cellcolor{LightCyan}$\hspace{-3mm}\begin{array}{c}\scriptscriptstyle{-0.08}\\ \scriptscriptstyle{(-0.22,0.05)}\end{array}$ & $\hspace{-3mm}\begin{array}{c}\scriptscriptstyle{-0.26}\\ \scriptscriptstyle{(-0.44,-0.08)}\end{array}$ & $\hspace{-3mm}\begin{array}{c}\scriptscriptstyle{-0.28}\\ \scriptscriptstyle{(-0.45,-0.09)}\end{array}$\tabularnewline \hline
$A_{03}=0$ &\cellcolor{LightCyan}$\hspace{-3mm}\begin{array}{c}\scriptscriptstyle{1.54}\\ \scriptscriptstyle{(-2.75,5.84)}\end{array}$ &\cellcolor{LightCyan}$\hspace{-3mm}\begin{array}{c}\scriptscriptstyle{3.25}\\ \scriptscriptstyle{(-1.23,8.07)}\end{array}$ &\cellcolor{LightCyan}$\hspace{-3mm}\begin{array}{c}\scriptscriptstyle{3.26}\\ \scriptscriptstyle{(-1.07,7.77)}\end{array}$ &\cellcolor{LightCyan}$\hspace{-3mm}\begin{array}{c}\scriptscriptstyle{0.27}\\ \scriptscriptstyle{(-0.38,0.93)}\end{array}$ &\cellcolor{LightCyan}$\hspace{-3mm}\begin{array}{c}\scriptscriptstyle{0.67}\\ \scriptscriptstyle{(-0.34,1.67)}\end{array}$ &\cellcolor{LightCyan}$\hspace{-3mm}\begin{array}{c}\scriptscriptstyle{0.59}\\ \scriptscriptstyle{(-0.41,1.59)}\end{array}$ &\cellcolor{LightCyan}$\hspace{-3mm}\begin{array}{c}\scriptscriptstyle{-0.01}\\ \scriptscriptstyle{(-0.08,0.06)}\end{array}$ &\cellcolor{LightCyan}$\hspace{-3mm}\begin{array}{c}\scriptscriptstyle{0.02}\\ \scriptscriptstyle{(-0.06,0.1)}\end{array}$ &\cellcolor{LightCyan}$\hspace{-3mm}\begin{array}{c}\scriptscriptstyle{0.02}\\ \scriptscriptstyle{(-0.06,0.09)}\end{array}$\tabularnewline \hline
$A_{04}=0$ &\cellcolor{LightCyan}$\hspace{-3mm}\begin{array}{c}\scriptscriptstyle{-3.51}\\ \scriptscriptstyle{(-8.17,1.13)}\end{array}$ &\cellcolor{LightCyan}$\hspace{-3mm}\begin{array}{c}\scriptscriptstyle{-2.53}\\ \scriptscriptstyle{(-7.76,2.61)}\end{array}$ &\cellcolor{LightCyan}$\hspace{-3mm}\begin{array}{c}\scriptscriptstyle{-2.53}\\ \scriptscriptstyle{(-7.61,2.41)}\end{array}$ &\cellcolor{LightCyan}$\hspace{-3mm}\begin{array}{c}\scriptscriptstyle{-0.96}\\ \scriptscriptstyle{(-2.09,0.2)}\end{array}$ &\cellcolor{LightCyan}$\hspace{-3mm}\begin{array}{c}\scriptscriptstyle{-1.6}\\ \scriptscriptstyle{(-3.35,0.18)}\end{array}$ &\cellcolor{LightCyan}$\hspace{-3mm}\begin{array}{c}\scriptscriptstyle{-1.41}\\ \scriptscriptstyle{(-3.14,0.27)}\end{array}$ &\cellcolor{LightCyan}$\hspace{-3mm}\begin{array}{c}\scriptscriptstyle{-0.01}\\ \scriptscriptstyle{(-0.04,0.01)}\end{array}$ & $\hspace{-3mm}\begin{array}{c}\scriptscriptstyle{-0.03}\\ \scriptscriptstyle{(-0.05,0)}\end{array}$ & $\hspace{-3mm}\begin{array}{c}\scriptscriptstyle{-0.02}\\ \scriptscriptstyle{(-0.05,0)}\end{array}$\tabularnewline \hline
$A_{05}=0$ &\cellcolor{LightCyan}$\hspace{-3mm}\begin{array}{c}\scriptscriptstyle{0.32}\\ \scriptscriptstyle{(-2.19,2.81)}\end{array}$ &\cellcolor{LightCyan}$\hspace{-3mm}\begin{array}{c}\scriptscriptstyle{1.85}\\ \scriptscriptstyle{(-1.6,5.2)}\end{array}$ &\cellcolor{LightCyan}$\hspace{-3mm}\begin{array}{c}\scriptscriptstyle{1.85}\\ \scriptscriptstyle{(-1.43,5.11)}\end{array}$ &\cellcolor{LightCyan}$\hspace{-3mm}\begin{array}{c}\scriptscriptstyle{0.66}\\ \scriptscriptstyle{(-0.11,1.43)}\end{array}$ & $\hspace{-3mm}\begin{array}{c}\scriptscriptstyle{1.18}\\ \scriptscriptstyle{(0.03,2.32)}\end{array}$ &\cellcolor{LightCyan}$\hspace{-3mm}\begin{array}{c}\scriptscriptstyle{1.12}\\ \scriptscriptstyle{(-0.02,2.25)}\end{array}$ &\cellcolor{LightCyan}$\hspace{-3mm}\begin{array}{c}\scriptscriptstyle{0}\\ \scriptscriptstyle{(-0.06,0.07)}\end{array}$ &\cellcolor{LightCyan}$\hspace{-3mm}\begin{array}{c}\scriptscriptstyle{0.05}\\ \scriptscriptstyle{(-0.04,0.13)}\end{array}$ &\cellcolor{LightCyan}$\hspace{-3mm}\begin{array}{c}\scriptscriptstyle{0.05}\\ \scriptscriptstyle{(-0.04,0.13)}\end{array}$\tabularnewline \hline
$A_{06}=-3$ & $\hspace{-3mm}\begin{array}{c}\scriptscriptstyle{-0.88}\\ \scriptscriptstyle{(-2.96,1.24)}\end{array}$ &\cellcolor{LightCyan}$\hspace{-3mm}\begin{array}{c}\scriptscriptstyle{-4.06}\\ \scriptscriptstyle{(-6.63,-1.76)}\end{array}$ &\cellcolor{LightCyan}$\hspace{-3mm}\begin{array}{c}\scriptscriptstyle{-4.02}\\ \scriptscriptstyle{(-6.5,-1.79)}\end{array}$ & $\hspace{-3mm}\begin{array}{c}\scriptscriptstyle{-1.8}\\ \scriptscriptstyle{(-2,-1.6)}\end{array}$ &\cellcolor{LightCyan}$\hspace{-3mm}\begin{array}{c}\scriptscriptstyle{-3.19}\\ \scriptscriptstyle{(-3.44,-2.94)}\end{array}$ &\cellcolor{LightCyan}$\hspace{-3mm}\begin{array}{c}\scriptscriptstyle{-3.06}\\ \scriptscriptstyle{(-3.3,-2.83)}\end{array}$ & $\hspace{-3mm}\begin{array}{c}\scriptscriptstyle{-1.71}\\ \scriptscriptstyle{(-1.77,-1.65)}\end{array}$ &\cellcolor{LightCyan}$\hspace{-3mm}\begin{array}{c}\scriptscriptstyle{-3.01}\\ \scriptscriptstyle{(-3.08,-2.94)}\end{array}$ & $\hspace{-3mm}\begin{array}{c}\scriptscriptstyle{-2.92}\\ \scriptscriptstyle{(-2.98,-2.85)}\end{array}$\tabularnewline \hline
$A_{07}=0$ &\cellcolor{LightCyan}$\hspace{-3mm}\begin{array}{c}\scriptscriptstyle{0.58}\\ \scriptscriptstyle{(-1.87,3.06)}\end{array}$ &\cellcolor{LightCyan}$\hspace{-3mm}\begin{array}{c}\scriptscriptstyle{0.44}\\ \scriptscriptstyle{(-2.27,3.17)}\end{array}$ &\cellcolor{LightCyan}$\hspace{-3mm}\begin{array}{c}\scriptscriptstyle{0.53}\\ \scriptscriptstyle{(-2.05,3.23)}\end{array}$ &\cellcolor{LightCyan}$\hspace{-3mm}\begin{array}{c}\scriptscriptstyle{0.05}\\ \scriptscriptstyle{(-0.49,0.58)}\end{array}$ &\cellcolor{LightCyan}$\hspace{-3mm}\begin{array}{c}\scriptscriptstyle{0.43}\\ \scriptscriptstyle{(-0.37,1.27)}\end{array}$ &\cellcolor{LightCyan}$\hspace{-3mm}\begin{array}{c}\scriptscriptstyle{0.39}\\ \scriptscriptstyle{(-0.42,1.19)}\end{array}$ &\cellcolor{LightCyan}$\hspace{-3mm}\begin{array}{c}\scriptscriptstyle{-0.02}\\ \scriptscriptstyle{(-0.07,0.04)}\end{array}$ &\cellcolor{LightCyan}$\hspace{-3mm}\begin{array}{c}\scriptscriptstyle{0.03}\\ \scriptscriptstyle{(-0.04,0.09)}\end{array}$ &\cellcolor{LightCyan}$\hspace{-3mm}\begin{array}{c}\scriptscriptstyle{0.03}\\ \scriptscriptstyle{(-0.03,0.09)}\end{array}$\tabularnewline \hline
$A_{08}=0$ &\cellcolor{LightCyan}$\hspace{-3mm}\begin{array}{c}\scriptscriptstyle{-0.83}\\ \scriptscriptstyle{(-3,1.31)}\end{array}$ &\cellcolor{LightCyan}$\hspace{-3mm}\begin{array}{c}\scriptscriptstyle{-2.22}\\ \scriptscriptstyle{(-4.98,0.62)}\end{array}$ &\cellcolor{LightCyan}$\hspace{-3mm}\begin{array}{c}\scriptscriptstyle{-2.12}\\ \scriptscriptstyle{(-4.84,0.65)}\end{array}$ &\cellcolor{LightCyan}$\hspace{-3mm}\begin{array}{c}\scriptscriptstyle{-0.48}\\ \scriptscriptstyle{(-1.03,0.07)}\end{array}$ & $\hspace{-3mm}\begin{array}{c}\scriptscriptstyle{-1}\\ \scriptscriptstyle{(-1.86,-0.14)}\end{array}$ & $\hspace{-3mm}\begin{array}{c}\scriptscriptstyle{-0.92}\\ \scriptscriptstyle{(-1.76,-0.09)}\end{array}$ &\cellcolor{LightCyan}$\hspace{-3mm}\begin{array}{c}\scriptscriptstyle{-0.02}\\ \scriptscriptstyle{(-0.08,0.03)}\end{array}$ &\cellcolor{LightCyan}$\hspace{-3mm}\begin{array}{c}\scriptscriptstyle{-0.02}\\ \scriptscriptstyle{(-0.09,0.05)}\end{array}$ &\cellcolor{LightCyan}$\hspace{-3mm}\begin{array}{c}\scriptscriptstyle{-0.02}\\ \scriptscriptstyle{(-0.09,0.05)}\end{array}$\tabularnewline \hline
$A_{09}=0$ &\cellcolor{LightCyan}$\hspace{-3mm}\begin{array}{c}\scriptscriptstyle{0.71}\\ \scriptscriptstyle{(-0.11,1.53)}\end{array}$ & $\hspace{-3mm}\begin{array}{c}\scriptscriptstyle{1.29}\\ \scriptscriptstyle{(0.06,2.61)}\end{array}$ & $\hspace{-3mm}\begin{array}{c}\scriptscriptstyle{1.33}\\ \scriptscriptstyle{(0.1,2.6)}\end{array}$ &\cellcolor{LightCyan}$\hspace{-3mm}\begin{array}{c}\scriptscriptstyle{0.33}\\ \scriptscriptstyle{(-0.02,0.67)}\end{array}$ & $\hspace{-3mm}\begin{array}{c}\scriptscriptstyle{0.7}\\ \scriptscriptstyle{(0.17,1.2)}\end{array}$ & $\hspace{-3mm}\begin{array}{c}\scriptscriptstyle{0.71}\\ \scriptscriptstyle{(0.19,1.2)}\end{array}$ & $\hspace{-3mm}\begin{array}{c}\scriptscriptstyle{0.08}\\ \scriptscriptstyle{(0.02,0.13)}\end{array}$ & $\hspace{-3mm}\begin{array}{c}\scriptscriptstyle{0.14}\\ \scriptscriptstyle{(0.05,0.23)}\end{array}$ & $\hspace{-3mm}\begin{array}{c}\scriptscriptstyle{0.15}\\ \scriptscriptstyle{(0.06,0.23)}\end{array}$\tabularnewline \hline
$A_{10}=0$ &\cellcolor{LightCyan}$\hspace{-3mm}\begin{array}{c}\scriptscriptstyle{-0.38}\\ \scriptscriptstyle{(-5.19,4.24)}\end{array}$ &\cellcolor{LightCyan}$\hspace{-3mm}\begin{array}{c}\scriptscriptstyle{-4.63}\\ \scriptscriptstyle{(-10.46,0.84)}\end{array}$ &\cellcolor{LightCyan}$\hspace{-3mm}\begin{array}{c}\scriptscriptstyle{-4.46}\\ \scriptscriptstyle{(-10.07,0.86)}\end{array}$ &\cellcolor{LightCyan}$\hspace{-3mm}\begin{array}{c}\scriptscriptstyle{-0.31}\\ \scriptscriptstyle{(-1.22,0.61)}\end{array}$ &\cellcolor{LightCyan}$\hspace{-3mm}\begin{array}{c}\scriptscriptstyle{0.66}\\ \scriptscriptstyle{(-0.55,1.87)}\end{array}$ &\cellcolor{LightCyan}$\hspace{-3mm}\begin{array}{c}\scriptscriptstyle{0.6}\\ \scriptscriptstyle{(-0.51,1.75)}\end{array}$ &\cellcolor{LightCyan}$\hspace{-3mm}\begin{array}{c}\scriptscriptstyle{0.02}\\ \scriptscriptstyle{(-0.13,0.17)}\end{array}$ &\cellcolor{LightCyan}$\hspace{-3mm}\begin{array}{c}\scriptscriptstyle{-0.02}\\ \scriptscriptstyle{(-0.2,0.16)}\end{array}$ &\cellcolor{LightCyan}$\hspace{-3mm}\begin{array}{c}\scriptscriptstyle{-0.01}\\ \scriptscriptstyle{(-0.2,0.16)}\end{array}$\tabularnewline \hline
$A_{11}=0$ &\cellcolor{LightCyan}$\hspace{-3mm}\begin{array}{c}\scriptscriptstyle{-2.93}\\ \scriptscriptstyle{(-8.11,2.15)}\end{array}$ &\cellcolor{LightCyan}$\hspace{-3mm}\begin{array}{c}\scriptscriptstyle{2.18}\\ \scriptscriptstyle{(-3.8,8.3)}\end{array}$ &\cellcolor{LightCyan}$\hspace{-3mm}\begin{array}{c}\scriptscriptstyle{2.24}\\ \scriptscriptstyle{(-3.53,8.27)}\end{array}$ & $\hspace{-3mm}\begin{array}{c}\scriptscriptstyle{0.88}\\ \scriptscriptstyle{(0.1,1.64)}\end{array}$ &\cellcolor{LightCyan}$\hspace{-3mm}\begin{array}{c}\scriptscriptstyle{0.1}\\ \scriptscriptstyle{(-0.89,1.09)}\end{array}$ &\cellcolor{LightCyan}$\hspace{-3mm}\begin{array}{c}\scriptscriptstyle{0.15}\\ \scriptscriptstyle{(-0.82,1.14)}\end{array}$ &\cellcolor{LightCyan}$\hspace{-3mm}\begin{array}{c}\scriptscriptstyle{-0.06}\\ \scriptscriptstyle{(-0.2,0.08)}\end{array}$ &\cellcolor{LightCyan}$\hspace{-3mm}\begin{array}{c}\scriptscriptstyle{0.08}\\ \scriptscriptstyle{(-0.09,0.26)}\end{array}$ &\cellcolor{LightCyan}$\hspace{-3mm}\begin{array}{c}\scriptscriptstyle{0.11}\\ \scriptscriptstyle{(-0.06,0.29)}\end{array}$\tabularnewline \hline
$A_{12}=5$ &\cellcolor{LightCyan}$\hspace{-3mm}\begin{array}{c}\scriptscriptstyle{5.96}\\ \scriptscriptstyle{(3.22,8.68)}\end{array}$ &\cellcolor{LightCyan}$\hspace{-3mm}\begin{array}{c}\scriptscriptstyle{8.28}\\ \scriptscriptstyle{(4.62,11.9)}\end{array}$ &\cellcolor{LightCyan}$\hspace{-3mm}\begin{array}{c}\scriptscriptstyle{7.82}\\ \scriptscriptstyle{(4.39,11.36)}\end{array}$ & $\hspace{-3mm}\begin{array}{c}\scriptscriptstyle{1.68}\\ \scriptscriptstyle{(0.72,2.65)}\end{array}$ &\cellcolor{LightCyan}$\hspace{-3mm}\begin{array}{c}\scriptscriptstyle{5.8}\\ \scriptscriptstyle{(4.52,7.07)}\end{array}$ &\cellcolor{LightCyan}$\hspace{-3mm}\begin{array}{c}\scriptscriptstyle{5.57}\\ \scriptscriptstyle{(4.36,6.81)}\end{array}$ & $\hspace{-3mm}\begin{array}{c}\scriptscriptstyle{2.88}\\ \scriptscriptstyle{(2.74,3.02)}\end{array}$ &\cellcolor{LightCyan}$\hspace{-3mm}\begin{array}{c}\scriptscriptstyle{5.05}\\ \scriptscriptstyle{(4.88,5.22)}\end{array}$ &\cellcolor{LightCyan}$\hspace{-3mm}\begin{array}{c}\scriptscriptstyle{4.89}\\ \scriptscriptstyle{(4.72,5.05)}\end{array}$\tabularnewline \hline
$A_{13}=0$ &\cellcolor{LightCyan}$\hspace{-3mm}\begin{array}{c}\scriptscriptstyle{2.68}\\ \scriptscriptstyle{(-1.67,6.93)}\end{array}$ &\cellcolor{LightCyan}$\hspace{-3mm}\begin{array}{c}\scriptscriptstyle{4.03}\\ \scriptscriptstyle{(-0.84,9.1)}\end{array}$ &\cellcolor{LightCyan}$\hspace{-3mm}\begin{array}{c}\scriptscriptstyle{4.08}\\ \scriptscriptstyle{(-0.63,8.96)}\end{array}$ & $\hspace{-3mm}\begin{array}{c}\scriptscriptstyle{-1.05}\\ \scriptscriptstyle{(-1.72,-0.38)}\end{array}$ &\cellcolor{LightCyan}$\hspace{-3mm}\begin{array}{c}\scriptscriptstyle{-0.83}\\ \scriptscriptstyle{(-1.76,0.11)}\end{array}$ &\cellcolor{LightCyan}$\hspace{-3mm}\begin{array}{c}\scriptscriptstyle{-0.81}\\ \scriptscriptstyle{(-1.72,0.1)}\end{array}$ &\cellcolor{LightCyan}$\hspace{-3mm}\begin{array}{c}\scriptscriptstyle{-0.01}\\ \scriptscriptstyle{(-0.08,0.05)}\end{array}$ &\cellcolor{LightCyan}$\hspace{-3mm}\begin{array}{c}\scriptscriptstyle{-0.01}\\ \scriptscriptstyle{(-0.09,0.08)}\end{array}$ &\cellcolor{LightCyan}$\hspace{-3mm}\begin{array}{c}\scriptscriptstyle{-0.01}\\ \scriptscriptstyle{(-0.09,0.08)}\end{array}$\tabularnewline \hline
$A_{14}=0$ &\cellcolor{LightCyan}$\hspace{-3mm}\begin{array}{c}\scriptscriptstyle{-1.13}\\ \scriptscriptstyle{(-5.84,3.6)}\end{array}$ &\cellcolor{LightCyan}$\hspace{-3mm}\begin{array}{c}\scriptscriptstyle{0.2}\\ \scriptscriptstyle{(-5.79,6.51)}\end{array}$ &\cellcolor{LightCyan}$\hspace{-3mm}\begin{array}{c}\scriptscriptstyle{0.02}\\ \scriptscriptstyle{(-5.85,6.12)}\end{array}$ & $\hspace{-3mm}\begin{array}{c}\scriptscriptstyle{1.39}\\ \scriptscriptstyle{(0.24,2.55)}\end{array}$ &\cellcolor{LightCyan}$\hspace{-3mm}\begin{array}{c}\scriptscriptstyle{0.12}\\ \scriptscriptstyle{(-1.44,1.66)}\end{array}$ &\cellcolor{LightCyan}$\hspace{-3mm}\begin{array}{c}\scriptscriptstyle{0.07}\\ \scriptscriptstyle{(-1.43,1.61)}\end{array}$ &\cellcolor{LightCyan}$\hspace{-3mm}\begin{array}{c}\scriptscriptstyle{0.02}\\ \scriptscriptstyle{(-0.01,0.05)}\end{array}$ & $\hspace{-3mm}\begin{array}{c}\scriptscriptstyle{0.03}\\ \scriptscriptstyle{(0,0.06)}\end{array}$ & $\hspace{-3mm}\begin{array}{c}\scriptscriptstyle{0.03}\\ \scriptscriptstyle{(0,0.06)}\end{array}$\tabularnewline \hline
$A_{15}=0$ &\cellcolor{LightCyan}$\hspace{-3mm}\begin{array}{c}\scriptscriptstyle{0.42}\\ \scriptscriptstyle{(-2.07,2.94)}\end{array}$ &\cellcolor{LightCyan}$\hspace{-3mm}\begin{array}{c}\scriptscriptstyle{1.49}\\ \scriptscriptstyle{(-1.84,5)}\end{array}$ &\cellcolor{LightCyan}$\hspace{-3mm}\begin{array}{c}\scriptscriptstyle{1.59}\\ \scriptscriptstyle{(-1.64,4.85)}\end{array}$ &\cellcolor{LightCyan}$\hspace{-3mm}\begin{array}{c}\scriptscriptstyle{-0.22}\\ \scriptscriptstyle{(-1.01,0.56)}\end{array}$ &\cellcolor{LightCyan}$\hspace{-3mm}\begin{array}{c}\scriptscriptstyle{0.15}\\ \scriptscriptstyle{(-0.89,1.18)}\end{array}$ &\cellcolor{LightCyan}$\hspace{-3mm}\begin{array}{c}\scriptscriptstyle{0.2}\\ \scriptscriptstyle{(-0.8,1.19)}\end{array}$ &\cellcolor{LightCyan}$\hspace{-3mm}\begin{array}{c}\scriptscriptstyle{0}\\ \scriptscriptstyle{(-0.07,0.06)}\end{array}$ &\cellcolor{LightCyan}$\hspace{-3mm}\begin{array}{c}\scriptscriptstyle{0.01}\\ \scriptscriptstyle{(-0.07,0.08)}\end{array}$ &\cellcolor{LightCyan}$\hspace{-3mm}\begin{array}{c}\scriptscriptstyle{0}\\ \scriptscriptstyle{(-0.07,0.08)}\end{array}$\tabularnewline \hline
$A_{16}=0$ &\cellcolor{LightCyan}$\hspace{-3mm}\begin{array}{c}\scriptscriptstyle{-0.82}\\ \scriptscriptstyle{(-2.94,1.29)}\end{array}$ & $\hspace{-3mm}\begin{array}{c}\scriptscriptstyle{-3.07}\\ \scriptscriptstyle{(-5.88,-0.39)}\end{array}$ & $\hspace{-3mm}\begin{array}{c}\scriptscriptstyle{-3.14}\\ \scriptscriptstyle{(-5.86,-0.5)}\end{array}$ &\cellcolor{LightCyan}$\hspace{-3mm}\begin{array}{c}\scriptscriptstyle{-0.08}\\ \scriptscriptstyle{(-0.28,0.12)}\end{array}$ &\cellcolor{LightCyan}$\hspace{-3mm}\begin{array}{c}\scriptscriptstyle{-0.15}\\ \scriptscriptstyle{(-0.45,0.15)}\end{array}$ &\cellcolor{LightCyan}$\hspace{-3mm}\begin{array}{c}\scriptscriptstyle{-0.15}\\ \scriptscriptstyle{(-0.45,0.15)}\end{array}$ &\cellcolor{LightCyan}$\hspace{-3mm}\begin{array}{c}\scriptscriptstyle{-0.03}\\ \scriptscriptstyle{(-0.09,0.03)}\end{array}$ &\cellcolor{LightCyan}$\hspace{-3mm}\begin{array}{c}\scriptscriptstyle{-0.07}\\ \scriptscriptstyle{(-0.15,0.02)}\end{array}$ & $\hspace{-3mm}\begin{array}{c}\scriptscriptstyle{-0.08}\\ \scriptscriptstyle{(-0.17,0)}\end{array}$\tabularnewline \hline
$A_{17}=0$ &\cellcolor{LightCyan}$\hspace{-3mm}\begin{array}{c}\scriptscriptstyle{-0.46}\\ \scriptscriptstyle{(-2.95,2.04)}\end{array}$ &\cellcolor{LightCyan}$\hspace{-3mm}\begin{array}{c}\scriptscriptstyle{-1.37}\\ \scriptscriptstyle{(-4.65,1.87)}\end{array}$ &\cellcolor{LightCyan}$\hspace{-3mm}\begin{array}{c}\scriptscriptstyle{-1.17}\\ \scriptscriptstyle{(-4.45,1.91)}\end{array}$ & $\hspace{-3mm}\begin{array}{c}\scriptscriptstyle{-0.71}\\ \scriptscriptstyle{(-1.25,-0.17)}\end{array}$ &\cellcolor{LightCyan}$\hspace{-3mm}\begin{array}{c}\scriptscriptstyle{-0.59}\\ \scriptscriptstyle{(-1.36,0.19)}\end{array}$ &\cellcolor{LightCyan}$\hspace{-3mm}\begin{array}{c}\scriptscriptstyle{-0.57}\\ \scriptscriptstyle{(-1.31,0.17)}\end{array}$ &\cellcolor{LightCyan}$\hspace{-3mm}\begin{array}{c}\scriptscriptstyle{0.01}\\ \scriptscriptstyle{(-0.05,0.06)}\end{array}$ &\cellcolor{LightCyan}$\hspace{-3mm}\begin{array}{c}\scriptscriptstyle{0.05}\\ \scriptscriptstyle{(-0.02,0.12)}\end{array}$ &\cellcolor{LightCyan}$\hspace{-3mm}\begin{array}{c}\scriptscriptstyle{0.05}\\ \scriptscriptstyle{(-0.03,0.12)}\end{array}$\tabularnewline \hline
$A_{18}=0$ &\cellcolor{LightCyan}$\hspace{-3mm}\begin{array}{c}\scriptscriptstyle{0.66}\\ \scriptscriptstyle{(-1.5,2.86)}\end{array}$ &\cellcolor{LightCyan}$\hspace{-3mm}\begin{array}{c}\scriptscriptstyle{0.3}\\ \scriptscriptstyle{(-2.65,3.15)}\end{array}$ &\cellcolor{LightCyan}$\hspace{-3mm}\begin{array}{c}\scriptscriptstyle{-0.01}\\ \scriptscriptstyle{(-2.73,2.67)}\end{array}$ & $\hspace{-3mm}\begin{array}{c}\scriptscriptstyle{0.65}\\ \scriptscriptstyle{(0.11,1.22)}\end{array}$ &\cellcolor{LightCyan}$\hspace{-3mm}\begin{array}{c}\scriptscriptstyle{0.24}\\ \scriptscriptstyle{(-0.51,0.99)}\end{array}$ &\cellcolor{LightCyan}$\hspace{-3mm}\begin{array}{c}\scriptscriptstyle{0.18}\\ \scriptscriptstyle{(-0.54,0.92)}\end{array}$ & $\hspace{-3mm}\begin{array}{c}\scriptscriptstyle{0.06}\\ \scriptscriptstyle{(0.01,0.12)}\end{array}$ &\cellcolor{LightCyan}$\hspace{-3mm}\begin{array}{c}\scriptscriptstyle{0.02}\\ \scriptscriptstyle{(-0.04,0.08)}\end{array}$ &\cellcolor{LightCyan}$\hspace{-3mm}\begin{array}{c}\scriptscriptstyle{0.02}\\ \scriptscriptstyle{(-0.04,0.08)}\end{array}$\tabularnewline \hline
$A_{19}=-3$ &\cellcolor{LightCyan}$\hspace{-3mm}\begin{array}{c}\scriptscriptstyle{-2.61}\\ \scriptscriptstyle{(-3.44,-1.78)}\end{array}$ & $\hspace{-3mm}\begin{array}{c}\scriptscriptstyle{-4.44}\\ \scriptscriptstyle{(-5.58,-3.38)}\end{array}$ & $\hspace{-3mm}\begin{array}{c}\scriptscriptstyle{-4.1}\\ \scriptscriptstyle{(-5.16,-3.13)}\end{array}$ & $\hspace{-3mm}\begin{array}{c}\scriptscriptstyle{-1.46}\\ \scriptscriptstyle{(-1.8,-1.12)}\end{array}$ &\cellcolor{LightCyan}$\hspace{-3mm}\begin{array}{c}\scriptscriptstyle{-3.09}\\ \scriptscriptstyle{(-3.53,-2.66)}\end{array}$ &\cellcolor{LightCyan}$\hspace{-3mm}\begin{array}{c}\scriptscriptstyle{-2.94}\\ \scriptscriptstyle{(-3.36,-2.55)}\end{array}$ & $\hspace{-3mm}\begin{array}{c}\scriptscriptstyle{-1.75}\\ \scriptscriptstyle{(-1.81,-1.69)}\end{array}$ &\cellcolor{LightCyan}$\hspace{-3mm}\begin{array}{c}\scriptscriptstyle{-3.06}\\ \scriptscriptstyle{(-3.14,-2.99)}\end{array}$ &\cellcolor{LightCyan}$\hspace{-3mm}\begin{array}{c}\scriptscriptstyle{-2.97}\\ \scriptscriptstyle{(-3.04,-2.9)}\end{array}$\tabularnewline \hline
 &\cellcolor{Violet} 8.48 & \cellcolor{Violet} 11.06 & \cellcolor{Violet} 10.32 & \cellcolor{Violet} 1.19 & \cellcolor{Violet} 0.76 & \cellcolor{Violet} 0.68 & \cellcolor{Violet} 0.43 & \cellcolor{Violet} 0.01 & \cellcolor{Violet} 0.01\tabularnewline 
\end{tabular}

}
\caption{Drift parameter estimates for a two dimensional cubic model with arbitrary diffusion function. On the left is the true value of the
parameter. The data used is the same as that of Table \ref{table:driftinfer} sampled at the $\Delta=0.1$ interval. In this case data is imputed
to obtain the intervals $\Delta=\{0.01,0.001\}$.
The bottom of the table shows the Posterior Expected Loss in each case.}
\label{table:driftinfer2}
\end{table}

Our aim is to infer models that can be used for prediction. This can be
problematic when dealing with non-linear models as some (generally unknown)
regions of the parameter space will give solutions that explode to infinity
with probability 1. This is a particular problem when, as exemplified by Table
\ref{table:driftinfer}, large amounts of data are needed to regain the true
values.

\begin{figure}
\begin{center}
  \epsfig{file=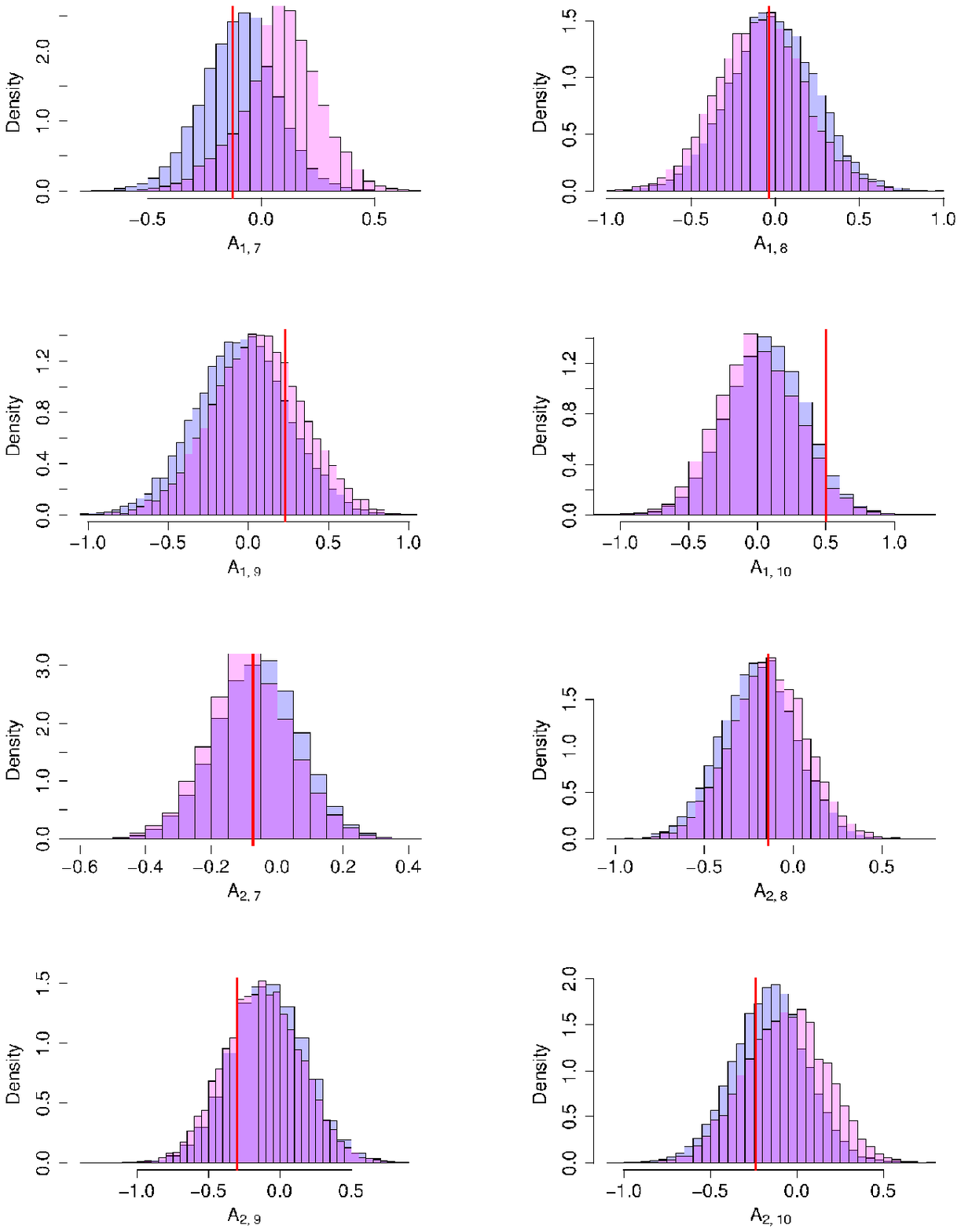,angle=0,scale=0.8}
 \end{center}
\caption{Marginal distributions of the cubic parameters inferred from a data set
  with $N=1,000$ observations at interval $\Delta=0.1$. The blue histogram
  shows the parameters that gave stable solutions to the SDE, while the red is
  for those that gave unstable solutions. The true values are given by the red
  lines.}
 \label{fig:instability}
\end{figure}

To demonstrate this problem we performed an inference on a two dimensional
cubic model using $N=1,000$ observations at $\Delta=0.1$. For each inferred
parameter value we then simulated the solution for $T=100$. After this time we
recorded whether the solution retained finite values or had exploded. The
marginal posterior distributions of the cubic parameters are plotted in Fig.
\ref{fig:instability}. Each plot shows two histograms: one in blue records the
distribution of stable parameter values and in red are those that
exploded. Notice that, when looking at the marginal distributions, the stable
and unstable regions largely overlap; it is difficult to separate the two
regions. In this case $40\%$ of values were unstable. Tests (not shown)
indicate that this is an even bigger problem in higher dimensions. Therefore,
it is essential to use constraints on the parameter space to enable
only physically meaningful solutions. The necessary conditions have
been derived in section \ref{Sec:2}.

Thus, as shown above, when updating the drift parameters we ensure that $\vect{M}$ is
negative definite. In practice it is sufficient to check only whether the symmetric
part $(\vect{M}+\vect{M}^T)/2$ is negative-definite. In the next
section we will develop a systematic way of sampling negative definite
matrices.

\subsection{Sampling Negative Definite Matrices \label{sec:negdefmats}}

To sample negative definite matrices we use the Component Wise
algorithm \cite{Peavoy:2013}. Here we sample the density of a $n\times n$ matrix $\vect{M}$,
with normally distributed components, subject to the constraint that
it is negative definite. This algorithm updates $\vect{M}$ component wise and
is based on the following property: a $n\times n$ matrix is negative
definite if and only if all $k\leq n$ leading principal minors obey
$|\vect{M}^{(k)}|(-1)^k>0$. The $k$th principal minor is the determinant of
the upper left $k\times k$ sub-matrix. Consider the parameters along
the main diagonal. As they only enter $\vect{M}$ once, each will have an
associated upper bound. The Algorithm  \ref{alg:diagonal} works by
calculating the upper bound associated with the constraints from each
principal minor. It does this to find the least upper bound and
thereby the truncation point of the normal distribution.

\begin{algorithm}
\caption{Sample parameters along diagonal}
\label{alg:diagonal}
\begin{algorithmic}
 \FOR{$i=1$ to $n$}
\STATE{$U_i$=0} 
\FOR{$j=i$ to $n$} 
\STATE{$x=-\left(\sum_{\mathtiny{\begin{array}{c} k\neq i\\k=1\end{array}}}^j(-1)^{i+k}M_{ik}|M^{(j)}_{\{-i\},\{-k\}}|\right)
/|M^{(j)}_{\{-i\},\{-i\}}|$} 
\ENDFOR
\IF{$x<U_i$}
\STATE{$U_i=x$}
\ENDIF
\STATE{$M_{ii}\sim \mathcal{N}_-(\mu_{i},U_i,\sigma^2_{i})$}
\ENDFOR
\end{algorithmic}
\end{algorithm}
Here, $\mathcal{N}_-(\mu,u,\sigma^2)$ is the right truncated normal
distribution with mean $\mu$, standard deviation $\sigma$ and upper bound
$u$. The off-diagonal parameters enter twice so there will be a quadratic
function determining their limits for each leading principal minor.  For
parameters in element $M^{(k)}_{ij}$ there will be an associated
quadratic relation
$a^{(k)}_{ij}M_{ij}^2+b^{(k)}_{ij}M_{ij}+c^{(k)}_{ij}=0$ where the
coefficients are functions of the other parameters. These coefficients are
found to be
\begin{subequations}
\begin{eqnarray} 
a^{(k)}_{ij}&=&-|M^{(k)}_{/\{i,j\},/\{i,j\}}|\\
b^{(k)}_{ij} &=& (-1)^{i+j}\sum_{\mathtiny{\begin{array}{c} k\neq
      i\\k=1\end{array}}}^{j-1}M_{jk}(-1)^{j-1+k}|M^{(k)}_{/\{i,j\},/\{j,k\}}|\\
& & + (-1)^{i+j}\sum_{\mathtiny{\begin{array}{c} k\neq
      i\\k=j+1\end{array}}}^{N}M_{jk}(-1)^{j+k}|M^{(k)}_{/\{i,j\},/\{j,k\}}|\\
&+&(-1)^{i+j}\sum_{\mathtiny{\begin{array}{c} k\neq
      j\\k=1\end{array}}}^{i-1}M_{ik}(-1)^{i-1+k}|M^{(k)}_{/\{i,j\},/\{i,k\}}|\\
& & + (-1)^{i+j}\sum_{\mathtiny{\begin{array}{c} k\neq
      j\\k=i+1\end{array}}}^{N}M_{ik}(-1)^{i+k}|M^{(k)}_{/\{i,j\},/\{i,k\}}|\\
c^{(k)}_{ij}&=& \sum_{\mathtiny{\begin{array}{c} k\neq
      j\\k=1\end{array}}}^{N}M_{ik}(-1)^{i+k}\left(\sum_{\mathtiny{\begin{array}{c}
      l\neq i\\l=1\end{array}}}^{k-1}
M_{jl}(-1)^{j-1+l}|M^{(k)}_{/\{i,j\},/\{l,k\}}| + \right. \\
& & \left. \sum_{\mathtiny{\begin{array}{c} l\neq i\\l=k+1\end{array}}}^{N}
M_{jl}(-1)^{j+l}|M^{(k)}_{/\{i,j\},/\{l,k\}}| \right),
\label{eq:offdiagcoefs}
\end{eqnarray}
\end{subequations}
where $|M^{(k)}_{/\{i,j\},/\{l,k\}}|$ represents the $k$th principal minor
with rows $i$ and $j$ and columns $l$ and $k$ removed. For each component
$M_{ij}$ this quadratic form can be solved to give upper and lower bounds on the
parameter. The matrix $\vect{M}$ can be cycled through updating each parameter in
turn. Algorithm \ref{alg:offdiagonal} describes the sampling of off-diagonal
elements using the coefficients in Eq. (\ref{eq:offdiagcoefs}). Here, the
notation, $\mathcal{N}_-^+(\mu,u^-,u^+,\sigma^2)$ refers to the doubly
truncated normal distribution with mean $\mu$, left truncation $u^-$, right
truncation $u^+$ and standard deviation $\sigma$.

\begin{algorithm}
\caption{Sample parameters off diagonal}
\label{alg:offdiagonal}
\begin{algorithmic}
 \FOR{$i=1$ to $n$}
\FOR{$j=i+1$ to $n$}
\STATE{$u^+=\infty$\\ $u^-=-\infty$}  
\FOR{$k=j$ to $n$}
\STATE{Calculate $a^{(k)}_{ij}$, $b^{(k)}_{ij}$ and $c^{(k)}_{ij}$ and solve $a^{(k)}_{ij}x^2+b^{(k)}_{ij}x+c^{(k)}_{ij}=0$.\\ 
Set $\mbox{mn }=\min(x_1,x_2)$ and $\mbox{mx }=\max(x_1,x_2)$} 
\ENDFOR
\IF{$mx<u^+$}
\STATE{$u^+=mx$}
\ENDIF
\IF{$mn>u^-$}
\STATE{$u^-=mn$}
\ENDIF
\STATE{$M_{ij}\sim \mathcal{N}_-^+(\mu_{ij},u^-,u^+,\sigma^2_{ij})$}
\ENDFOR
\ENDFOR
\end{algorithmic}
\end{algorithm}

To simulate from truncated normal distributions we are using the inverse
Cumulative Density Function (CDF) method. One simply calculates the
corresponding CDF of the lower and upper boundaries and then draws a uniform
random variable between these numbers. Inverting the CDF gives a random
variable from the Normal distribution restricted to this region.

For our problem we use the rejection sampler method proposed by
\cite{Robert:1995}. This method draws uncorrelated samples directly from the
target density. Rejection sampling from a distribution $h(x)$ is based on a
proposal distribution $g(x)$ such that $h(x)\leq Cg(x)$ holds for some
constant $C$ and all of the support of $h(x)$. For a one sided truncated
Normal the exponential distribution is a good proposal. First it is translated
to coincide with the truncation point, then the rate parameter is optimized in
order to closely match the tail of the Normal distribution.

\be g(z;\alpha,\mu^-)=\alpha\exp(-\alpha(z-\mu^-))\mathbb{I}_{z\geq \mu^-}\ee
The optimal value of $\alpha$ is calculated by maximizing the expected
acceptance probability and is shown to be 
\be \alpha^*(\mu^-)=\frac{\mu^-+\sqrt{(\mu^-)^2+4}}{2} \ee
More details are given in \cite{Robert:1995}.

We performed a numerical study to compare the standard
Normal and Exponential proposals. The efficiency of proposing $x$ from the
standard normal and then accepting if $x>\mu^-$ falls to approximately $0.023$
while for the optimized exponential proposal it is approximately $0.5$.

\begin{figure}
\begin{center}
  \epsfig{file=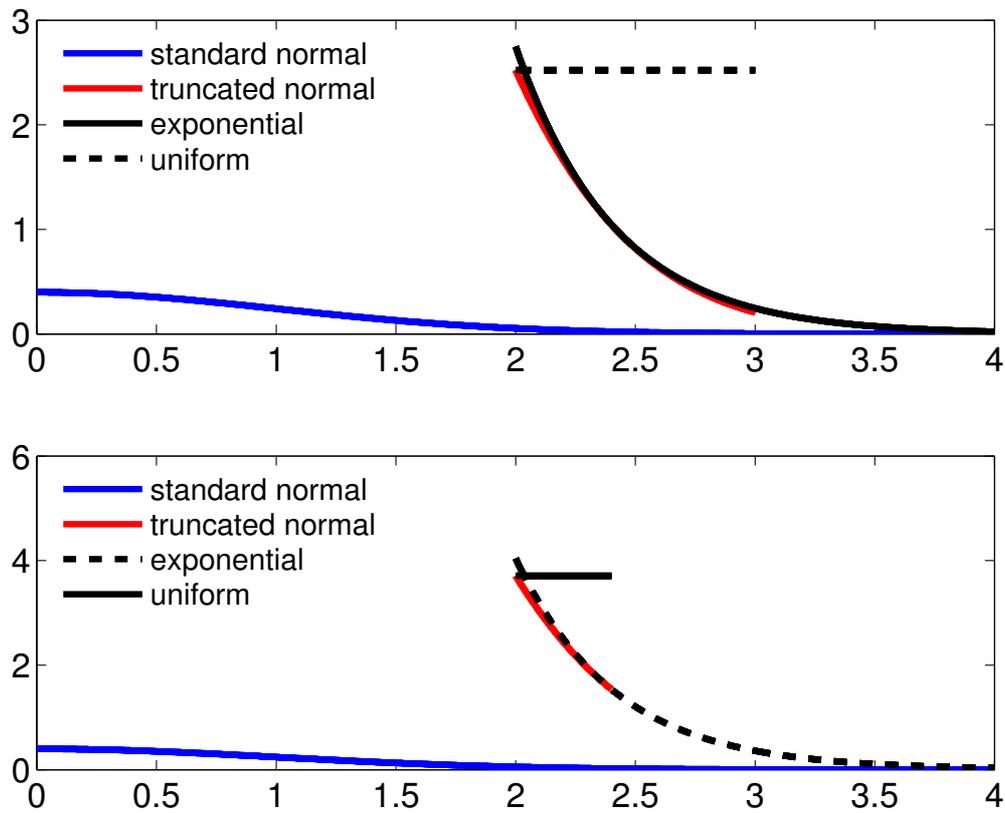,angle=0,scale=0.8}
\end{center}
\caption{Doubly truncated normal distribution. The top figure has $u^-=2$ and
  $u^+=3$ and is better approximated with the exponential distribution. The
  bottom figure has $u^-=2$ and $u^+=2.5$ and the uniform is more efficient.}
\label{fig:doubletrunc}
\end{figure}

For the doubly truncated Normal one uses either an exponential or uniform
distribution, as a proposal, depending upon the size of the truncated
region. If the following holds 
\[ u^+>u^-+\frac{2\sqrt{e}}{u^-+\sqrt{(u^-)^2+4}}\exp(\frac{(u^-)^2-u^-\sqrt{(u^-)^2+4}}{4})\]
then it can be shown that the exponential is more efficient, otherwise the
uniform is better \citep{Robert:1995}. Fig. \ref{fig:doubletrunc} shows both
the uniform and exponential approximations for both cases.

\begin{figure}[!t]
\begin{center}
\epsfig{file=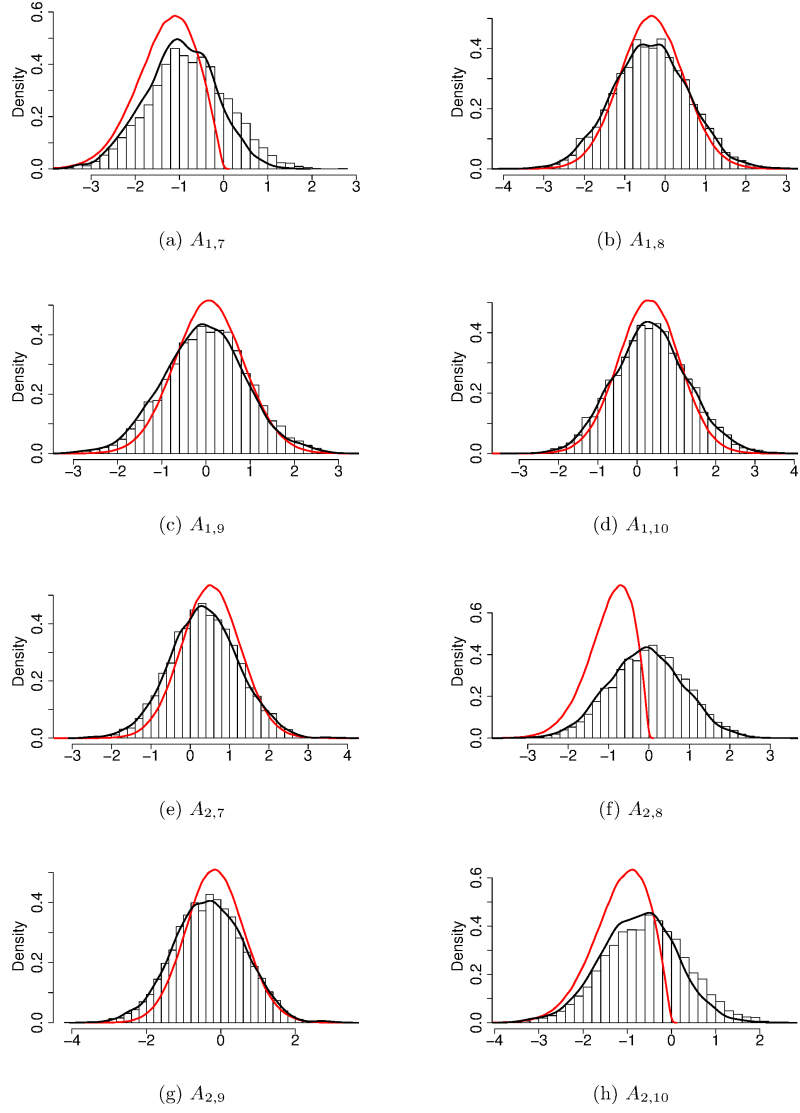,angle=0,scale=0.8}
\end{center}
\caption{Estimated posterior distributions for parameters from a two
  dimensional model of the form Eq. \ref{EQ:0} with N = 100 and ∆ = 0.1. The
  parameters, which are randomly generated, are written in the matrix notation
  introduced in Section \ref{sec:gibbs}. The histograms are the posterior
  distributions with uninformative prior, in red are the posterior
  distributions for parameters with stable SDEs and in black are the posterior
  distributions which include the stability matrix prior information derived
  in this chapter.}
\label{fig:posteriorscomponentwise}
\end{figure}

We use Algorithms \ref{alg:diagonal} and \ref{alg:offdiagonal}, along with the
methods of sampling truncated Normal variables, to sample the stability matrix
$\vect{M}$. We tested this
algorithm on a three dimensional model with $T=100$. In this case the
dimension of $\vect{M}$ is $n=D(D+1)/2=6$. Note
that this MCMC algorithm still mixes well. Fig.
\ref{fig:posteriorscomponentwise} compares the posterior distributions
estimated from the Component Wise Algorithm and the
standard Gibbs sampler. Notice the large differences between the distributions
in each case. Similar results (not shown) are obtained for the off diagonal
parameters using Algorithm \ref{alg:offdiagonal}.

\section{Results}
\label{Sec:4}

\subsection{Deterministic Double Well Potential Model\label{sec:lorenz}}

\begin{figure}[!t]
\begin{center}
\epsfig{file=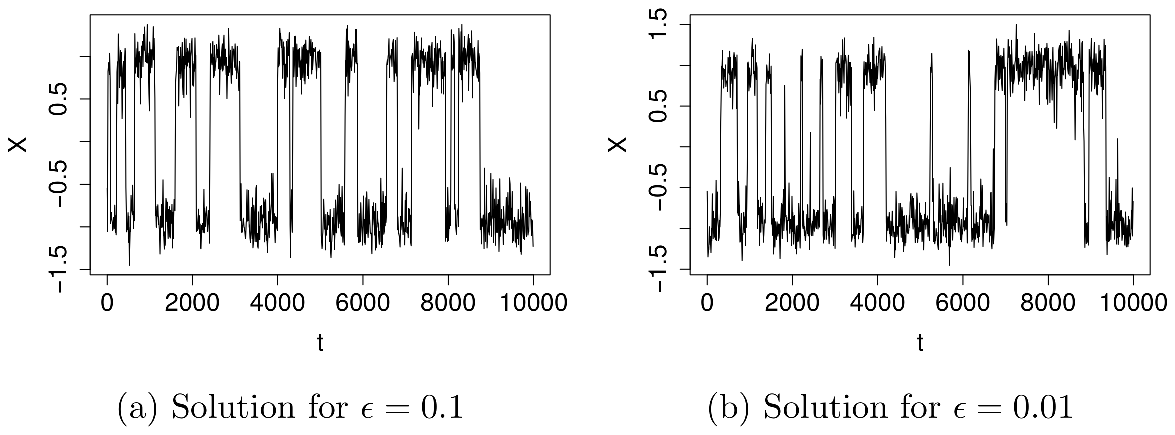,angle=0,scale=1}
\end{center}
\caption{Example path of x of the chaotic Lorenz system: Eq. (\ref{eq:lorenz})}
\label{fig12}
\end{figure}

\begin{figure}[!t]
\centering
\epsfig{file=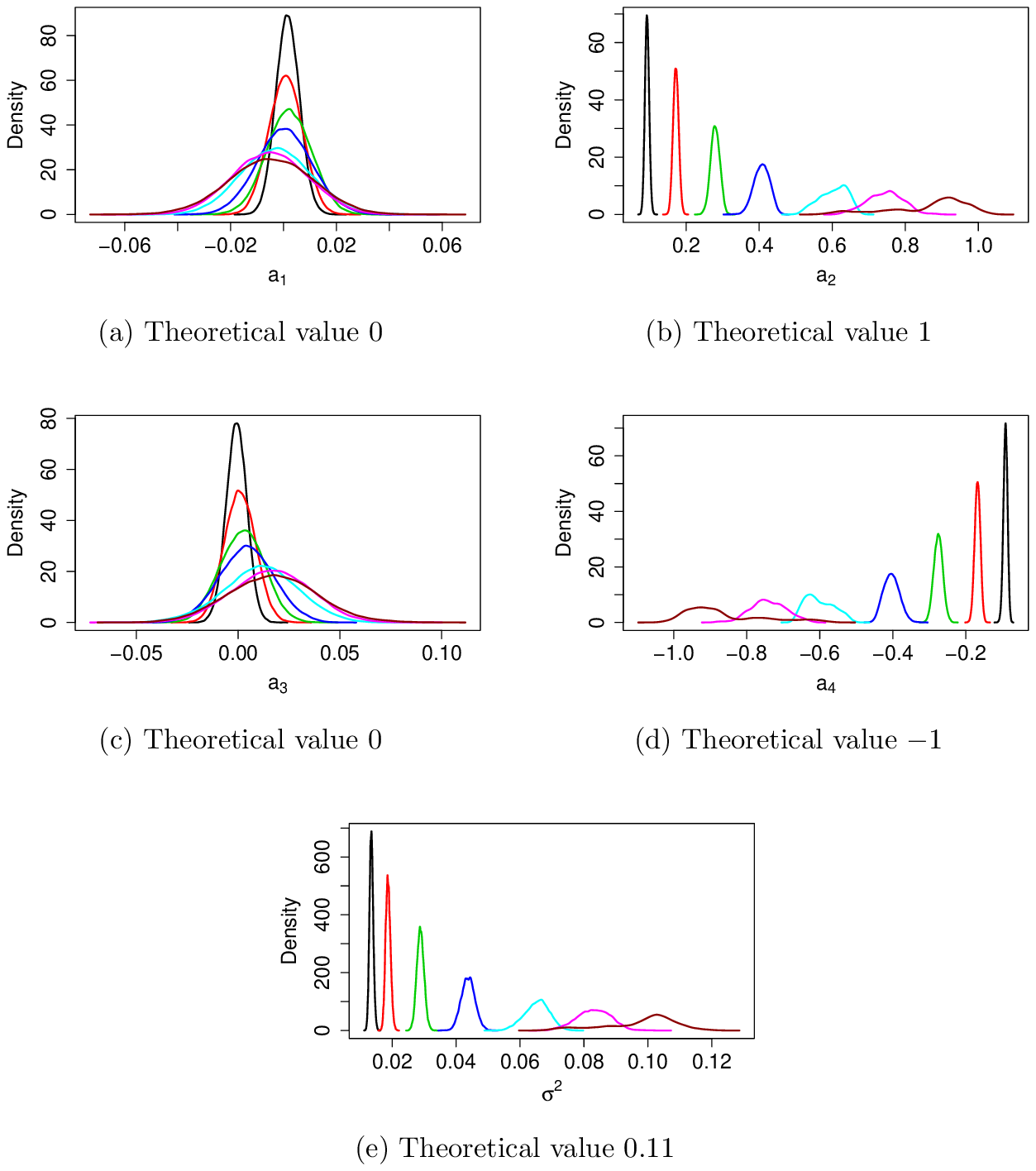,angle=0,scale=1}
\caption{Posterior distribution estimates from MCMC output applied to
  a sparse data set ($\Delta=10$). Different distributions correspond
  to increasing amounts of missing data. The distribution in brown,
  for $m=64$, agrees with the theoretical values predicted by the
  homogenization procedure. In the model simulation $ \epsilon = 0.01
  $ has been used. Black line: m=1, red line: m=2, green line: m=4,
  blue line: m=8, light blue line: m=16, magenta line: m=32, brown line: m=64.}
\label{fig:doublewellmcmc}
\end{figure}

The first conceptual climate model we consider is a cubic model coupled to the chaotic Lorenz
system \citep{Mitchell:2012}. It is fully deterministic and consists of a slow
variable which can be thought of as representing a climate process and three fast
variables which can be thought of as 
representing chaotic 
weather fluctuations. The slow variable moves inside a double well potential
and is perturbed by the chaotic Lorenz system, which acts effectively as
noise when $ \epsilon \rightarrow 0 $. The equations are as follows
\begin{subequations}
\begin{eqnarray}
\frac{dx}{dt}&=&x-x^3+\frac{4}{90\epsilon}y_2\\
\frac{dy_1}{dt}&=&\frac{10}{\epsilon^2}(y_2-y_1)\\
\frac{dy_2}{dt}&=&\frac{1}{\epsilon^2}(28y_1-y_2-y_1y_3)\\
\frac{dy_3}{dt}&=&\frac{1}{\epsilon^2}(y_1y_2-\frac{8}{3}y_3)\,.\label{eq:lorenz}
\end{eqnarray}
\end{subequations}
Sample paths are displayed in Fig. \ref{fig12}. We now fit a one dimensional
cubic SDE to the data. We just consider the general cubic form
\citep{Majda:2009}
\be
dX_t=(a_1+a_2X_t+a_3X_t^2+a_4X_t^3)dt+\sigma dW_t \label{eq:standard}
\ee and
estimate all of the parameters $\{a_1,a_2,a_3,a_4,\sigma\}$ from sparse
observations of the system: again using $\Delta=10.0$ and $N=1000$. To update
the drift parameters we use the Gibbs sampler of Section \ref{sec:gibbs}. The
estimated posterior distributions are shown in Fig.
\ref{fig:doublewellmcmc}. A lot of imputed data is needed before the estimates
start to converge towards the values predicted by homogenization but the
inference demonstrates that there is enough information in the sparse data set
if the likelihood is well approximated.

\begin{figure}[!t]
\begin{center}
\epsfig{file=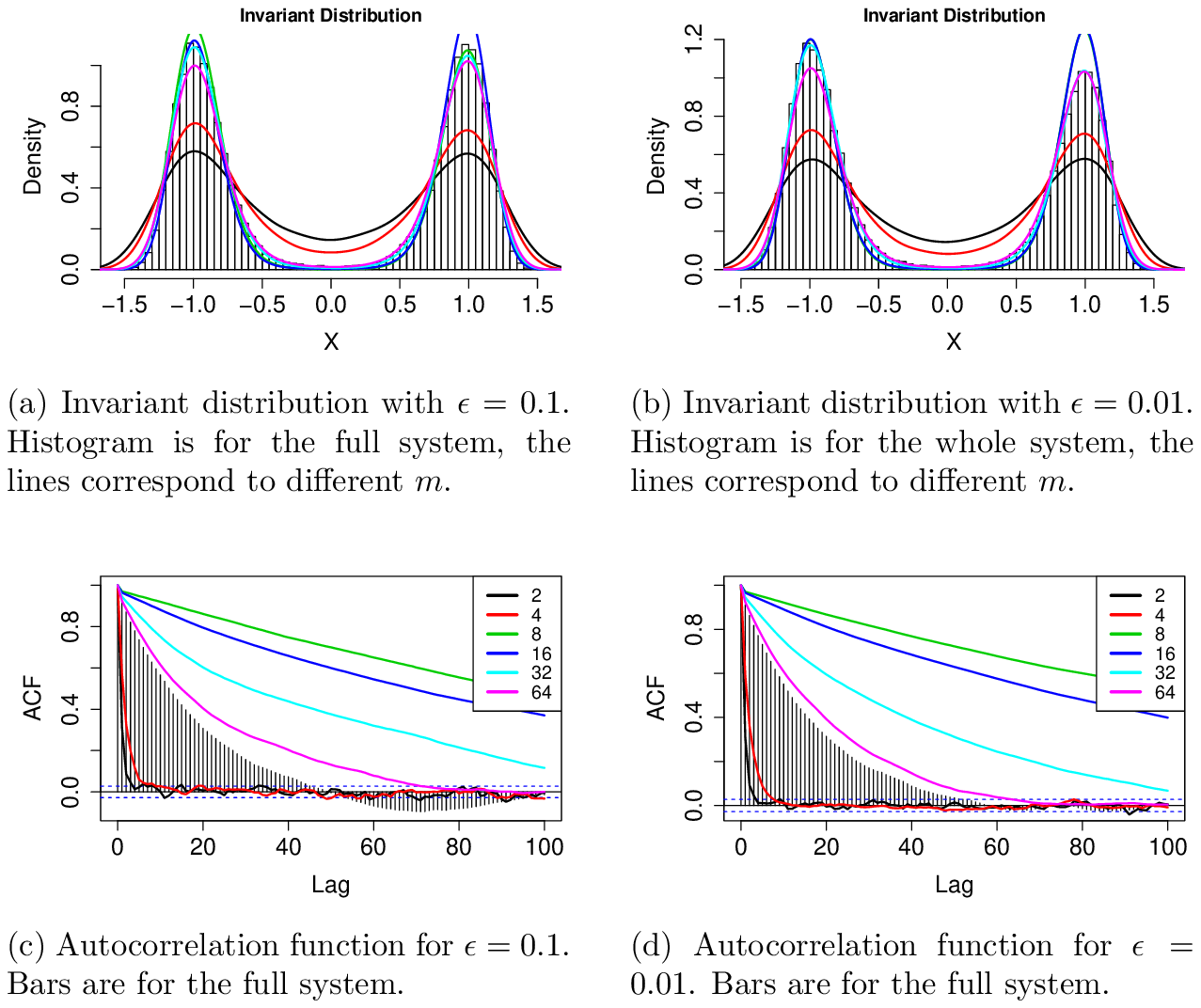,angle=0,scale=1}
\end{center}
\caption{Predictive statistics for the reduced double well model coupled to
  chaotic Lorenz system: Eq. (\ref{eq:lorenz}) for two values of
  $\epsilon$. In each plot the lines correspond to the inferred one
  dimensional model for different $m$.}
\label{fig:doublewellpred}
\end{figure}

Figure \ref{fig:doublewellpred} shows the predictive skill of the one
dimensional reduced model for $\sigma$ estimated for various $m$. We
use the empirical mean estimate for the parameter values. Figures
\ref{fig:doublewellpred}a and \ref{fig:doublewellpred}b show that the reduced
model can reproduce the double well distribution of the full model although the
separation of each well is underestimated for $m=2$ and $m=4$ due to the
larger noise. For $m\geq 8$ the model reproduces well the full models marginal
distribution for $x$. It is not clear whether there is much difference between
$m=8$ and $m=64$. However, observing Figures \ref{fig:doublewellpred}c and
\ref{fig:doublewellpred}d we see that the autocorrelation function for the
full model is much better approximated when $m=64$. This shows that
the ability of our framework to impute data is a powerful way of
deriving accurate reduced order models.

\subsection{Model Reduction for Triad Systems\label{sec:infertriad}}

Now we apply our model fitting procedure to a triad model with a high
dimensional deterministic system with two slow, climate variables coupled to
fast chaotic dynamics. The reduction strategy has two challenges: to
successfully approximate the deterministic variables by a stochastic process
and to be insensitive to a lack of time scale separation. 

\begin{figure}[!t]
\begin{center}
\epsfig{file=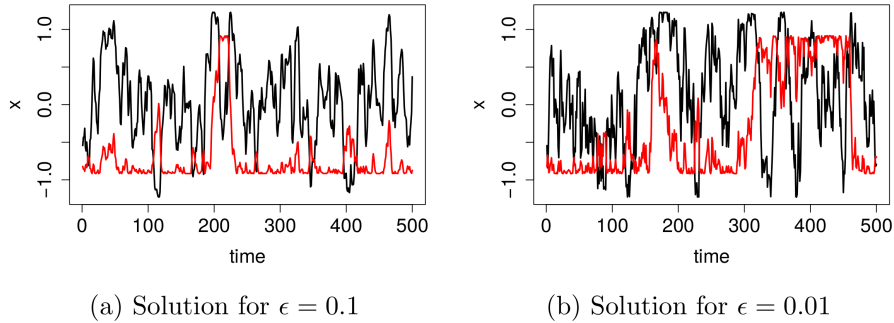,angle=0,scale=1}
\end{center}
\caption{Example path of $x_1$ (black line) and $x_2$ (red line) from the
  triad model: Eq. (\ref{eq:fullburgers})}
\label{fig13}
\end{figure}

The full system is given by
\begin{subequations}
\begin{eqnarray}
 \frac{dx_1}{dt} &=& \frac{b_1}{\epsilon}x_2y_1\\
\frac{dx_2}{dt} &=& \frac{b_2}{\epsilon}x_1y_1\\
\frac{dy_k}{dt} &=& \frac{b_3}{\epsilon}x_1x_2\delta_{1,k} -\mbox{Re} \frac{ik}{2\epsilon^2} \sum_{p+q+k=0}\hat{u}_p^*\hat{u}_q^*\\
\frac{dz_k}{dt} &=& -\mbox{Im}\frac{ik}{2\epsilon^2} \sum_{p+q+k=0}\hat{u}_p^*\hat{u}_q^*,
\end{eqnarray}
\label{eq:fullburgers}
\end{subequations}
where $ u_k = y_k + i z_k $. This system is stable provided that the
energy is conserved: $b_1+b_2+b_3=0$. We use the values
$\vect{b}=\{0.9,-0.5-0.4\}$ (our results are insensitive over a wide range of
parameter values) and we choose a cut off of
$\Lambda=50$. Sample paths are displayed in Fig. \ref{fig13}. We are
interested in eliminating $\vect{y}$ leaving equations for just $x_1$ and
$x_2$. The small parameter $\epsilon$ represents the time scales within the
system. The variables $\vect{y}$ have fastest time scale of order
$O(1/\epsilon^2)$ compared to $O(1/\epsilon)$ for $x_1$ and $x_2$. As
$\epsilon\rightarrow 0$ we can use the method of homogenization for SDEs to
eliminate the fast variables and this gives
\begin{subequations}
\begin{eqnarray}
dx_1(t)&=&\frac{b_1}{\gamma}(b_3x_2^2(t)+\frac{\sigma^2}{2\gamma}b_2)x_1(t)dt+\frac{\sigma}{\gamma}b_1x_2(t)dW_t\\
dx_2(t)&=&\frac{b_2}{\gamma}(b_3x_1^2(t)+\frac{\sigma^2}{2\gamma}b_1)x_2(t)dt+\frac{\sigma}{\gamma}b_2x_1(t)dW_t\,,
\end{eqnarray}
\label{eq:reducedburgers}
\end{subequations}
where unknown parameters $\sigma$ and $\gamma$ have been introduced. Here we
estimate them using the Algorithms \ref{alg:diffusionparams} and
\ref{alg:updatedata} from observations of the climate variables
alone. For convenience we consider the inference problem where the model
Eq. (\ref{eq:reducedburgers}) is driven by two independent Brownian motions.

\begin{figure}[!t]
\begin{center}
\epsfig{file=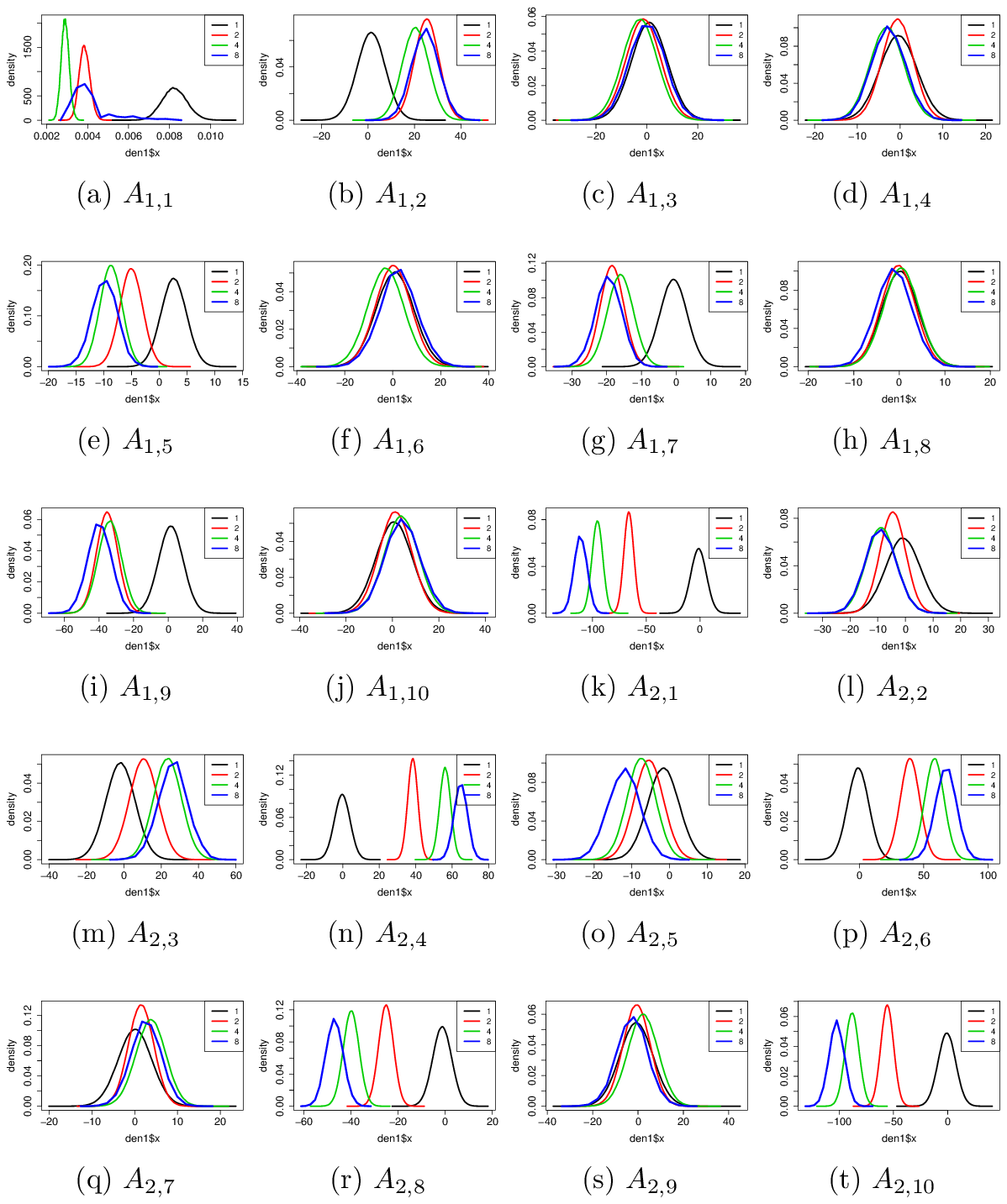,angle=0,scale=1}
\end{center}
\caption{Posterior estimates of drift parameters for two dimensional cubic
  model fitted to the triad-Burgers equation for $\epsilon=0.8$.}
\label{fig:empiricalburgers}
\end{figure}

Posterior estimates for the case $\epsilon=0.8$ are shown in
Fig. \ref{fig:empiricalburgers}. This value corresponds to a
moderately small, though realistic \cite{Franzke:2005}, amount of time
scale separation. As Fig. \ref{fig:empiricalburgers} demonstrates, increasing
the number of imputed data leads to a convergence and, thus, to an improvement
of the posterior estimates. 

\begin{figure}[!t]
\begin{center}
\epsfig{file=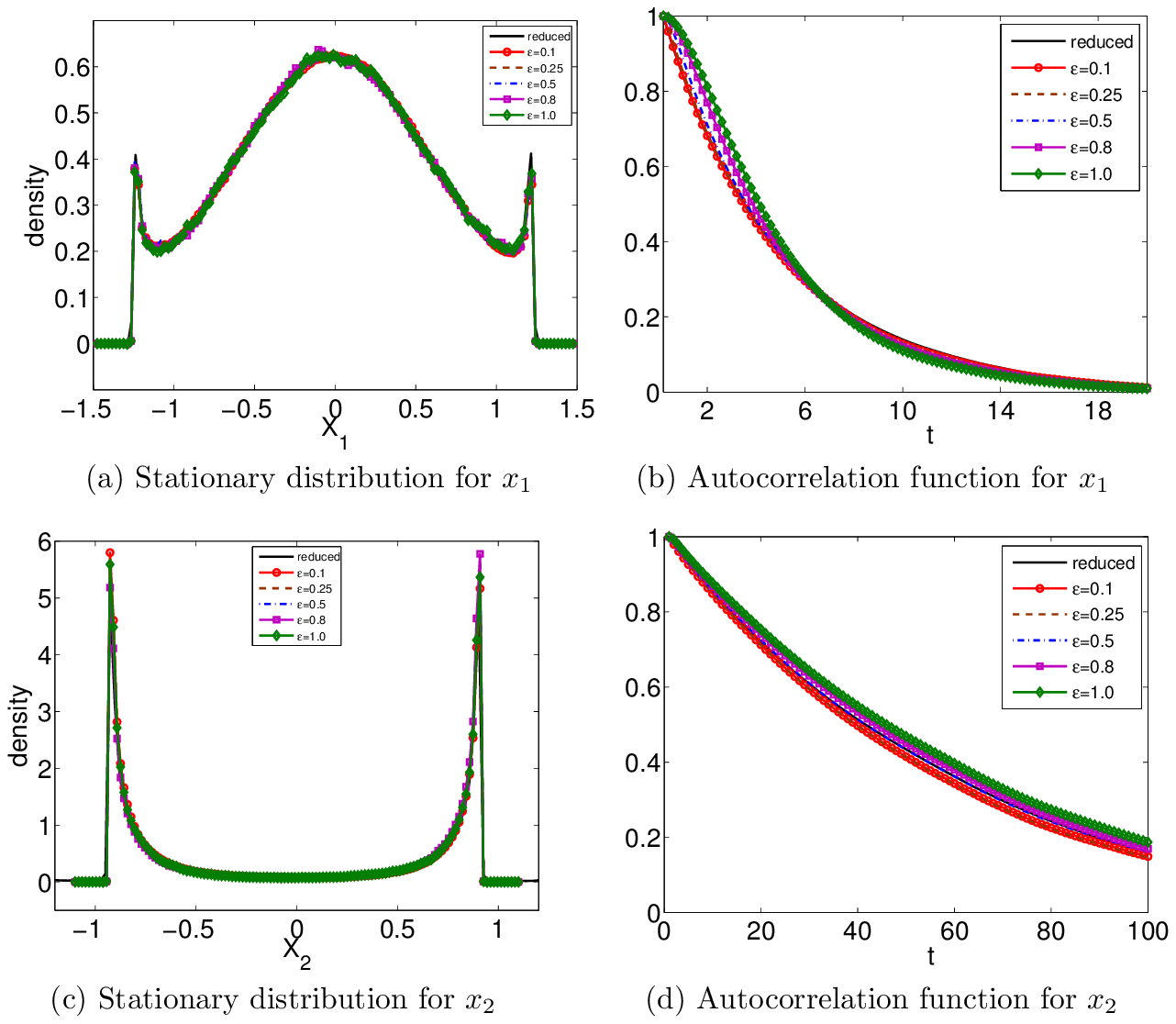,angle=0,scale=1.0}
\end{center}
\caption{Output statistics comparing the reduced model
  Eq. \ref{eq:reducedburgers}, with empirical models for various
  $\epsilon$ values with the full model.} 
\label{fig:reducedburgersestimates}
\end{figure}

We apply the inference to a data set with total time $T=500$ and observation
interval $\Delta=0.1$. We simulate the system for
$\epsilon=\{0.1,0.25,0.5,0.8,1.0\}$. Fig. \ref{fig:reducedburgersestimates}
shows the predictive probability densities and autocorrelation functions for
Eq. \ref{eq:reducedburgers}. In each case the data is simulated from the full
model Eq. \ref{eq:fullburgers}, then the parameters are estimated using the
reduced model Eq. \ref{eq:reducedburgers} and this reduced model is simulated
to calculate the predictive statistics. The posterior mean estimates were
computed for m = 16 missing data values (it was veriﬁed that the posteriors
for m = 16 and m $<$ 16 gave consistent estimates). The reduced model
Eq. \ref{eq:reducedburgers} with empirical parameter estimates is also
plotted. This model is referred to as the reduced model in Figure
\ref{fig:reducedburgersestimates}. The reduced model is able to
reproduce the non-trivial shape of the PDF very well. This suggests
tha reduced order models fitted from observed data can be used for
extreme value studies \cite{Franzke:2012}. 

The autocorrelation 
functions have been collapsed onto the reduced model by rescaling the output
interval of the prediction by their value of $\epsilon$; this has been done
for convenience of displaying the results. The data collapse is very good for
all model simulations with the models with $ \epsilon < 0.5 $ being closest to
the reduced model. This implies that the parameter estimates for each case are
partially compensating for the changing time scale separation. This provides
evidence for the potential of using reduced order modelling strategies even in
systems with only moderate time scale separation.

%

\section{Summary}
\label{Sec:6}
Here we developed a systematic Bayesian framework for the inference
of the parameter of SDEs constrained by the physics of the underlying
system. The physical constraints not only constrain the parameter space
but also enforce global stability of the reduced order models.

For climate models we derive a constraint based on energy
conservation which ensures global stability of the effective
SDE. This constraint takes the form of a negative definite matrix. We
then develop a new algorithm for the sampling of negative definite
matrices. We also develop a new algorithm for imputing data and show
that imputing data improves the accuracy considerly of the parameter
estimation. We demonstrated its power successfully on two conceptual
climate models.

While we focused on climate models in this study our method is general
enough that it can also be applied to other areas of fluid
dynamics. Furthermore, also many other physical systems observe
conservation laws and, thus, stability conditions can be derived which
will be useful for parameter estimation procedures.

{\bf Acknowledgments}
We thank three anonymous reviewers whose comments improved earlier versions
of this manuscript. This study received funding from the Natural Environment
Research Council, the Engineering and Physical Sciences Research Council and
the Deutsche Forschungsgemeinschaft (DFG) through the cluster of excellence
CliSAP.

\end{document}